\documentclass[12pt]{elsart}
\usepackage{color}

\usepackage{amsmath}
\usepackage{amsfonts}
\usepackage{mathrsfs}
\usepackage{graphicx}
\date{}

\newcommand{\nl}{\nonumber\\ }
\newcommand{\pd}{\partial}

\def\bc{\begin{center}}
\def\ec{\end{center}}
\def\om{\omega}
\def\be{\begin{eqnarray}}
\def\ee{\end{eqnarray}}
\def\prt{\partial}
\def\lsim{\stackrel{\scriptstyle <}{\phantom{}_{\sim}}}
\def\gsim{\stackrel{\scriptstyle >}{\phantom{}_{\sim}}}
\def\rmd{{\rm d}}

\begin{document}
\begin{frontmatter}

\title{Viscosity coefficients  for hadron and quark-gluon phases}
\author[JINR,Mold]{A.S.~Khvorostukhin},
\author[JINR,GSI]{V.D.~Toneev}
\author[MEPHI,GSI]{D.N. Voskresensky}
\address[JINR]{Joint Institute for Nuclear Research,
 141980 Dubna, Moscow Region, Russia}
\address[Mold]{ Institute of Applied Physics, Moldova Academy of Science,
MD-2028 Kishineu, Moldova}
\address[GSI]{GSI, Plankstra\ss{}e 1, D-64291 Darmstadt, Germany}
\address[MEPHI]{Moscow Engineering Physical Institute,\\ Kashirskoe
  Avenue 31, RU-115409 Moscow, Russia}
\maketitle{}
\begin{abstract}
The shear ($\eta$) and bulk ($\zeta$) viscosities are calculated
in a quasiparticle relaxation time approximation. The hadron phase
is described within the relativistic mean field based model with
scaled hadron masses and couplings. The quark phase is treated in
terms of the heavy quark bag model fitted to the lattice data. A
two-phase model allowing for the first order phase transition from
the hadron phase to the strongly coupled quark gluon plasma is
constructed by means of the Gibbs conditions.
Temperature and baryon density dependence of the calculated
viscosity-to-entropy ratios ($\eta/s$, $\zeta/s$) are analyzed and
compared with those obtained in other models. Special attention is
paid to the behavior of viscosity coefficients near the critical
temperature, from both hadron and quark-gluon side.  Effects of
resonance widths  on viscosities and viscosity-to-entropy ratios
 are  estimated.
\end{abstract}\end{frontmatter}

\section{Introduction}

The study of the transport properties of nonequilibrium  systems
not far from an equilibrium state has a very long story. Methods
for the calculation of  transport coefficients  were probed,
e.g., in description of nonrelativistic classical
gases~\cite{ChC70}, liquids and glasses~\cite{E83}, relativistic
gases~\cite{GLW80},  cold atomic gases \cite{Tschafer},
Fermi~\cite{BP78} and Bose~\cite{Khalatnikov} liquids. In the
past, transport coefficients for the nuclear matter were also
studied \cite{AW73,Gal79,Dan84,HMPS92,HM93}.   A knowledge of
various transport coefficients is required also in astrophysical
problems such as  the  entropy production in the universe,  the
electro-weak baryogenesis~\cite{W71},  for the description of
various phenomena in supernovas and neutron stars~\cite{Thompson}.
A recent paper \cite{Shlomo} considered shear viscosity effects in
excited atomic nuclei. With accessibility of heavy-ion collisions
the possibility for the creation of a new state of matter, the
quark-gluon plasma (QGP), has been offered. In this respect, there
appeared  growing interest in the calculation of the transport
coefficients of the
QGP~\cite{Tschafer,DG85,G85,HK85,BMPR90,Hsong}. Transport
coefficients in hot gauge theories were considered in Refs.
 \cite{AMY00,PSS01,M01,VB02,AM02,BLS05,D05,PC05}.
Attention has mostly been focused on ultrarelativistic
theories, where the scale is set exclusively by the temperature
$T$. These results are valid for $T$ much higher than the critical
temperature $T_c$ of the deconfinement phase transition.   Near
the critical temperature the highly nontrivial effects of the
strong coupling should be important.

 Recently, the
interest in the transport coefficient issue  has sharply been
increased in  heavy-ion collision physics. Large values of the
elliptic flow $v_2$ were observed at the Relativistic Heavy Ion
Collider~\cite{RHIC-v2}. This finding indicates that the created
QGP behaves as a nearly perfect fluid with a small value of the
shear viscosity-to-entropy density ratio, $\eta/s$, which was
confirmed by non-ideal hydrodynamic analysis of these
data~\cite{Tea03}. It was claimed~\cite{PSS01,PC05,Sh05} that a
new state produced at high temperatures is most likely not a
weakly interacting QGP, as it was originally assumed, but a
strongly interacting quark-gluon plasma (sQGP). This interest was
also supported by a new theoretical perspective, namely,  ${\cal
N}=4$ supersymmetric Yang-Mills gauge theory using the Anti de-Sitter
space/Conformal Field Theory (AdS/CFT) duality conjecture. Calculations
in this strongly coupled theory  closely related to QCD give for the
$\eta/s$ ratio~\cite{PSS01,BLS05,KSS03}
\begin{equation}
\frac{\eta}{s}=\frac{1}{4\pi} \left(1+\frac{135
\zeta(3)}{8(2\lambda)^{3/2}}+...\right).
\end{equation}
This result was confirmed in subsequent
investigations~\cite{HJM07,PW08,LH08}. After evaluating $\eta/s$
in several strongly coupled theories it was conjectured that
$\eta/s=\frac{1}{4\pi}$ is in fact a lower bound on the $\eta/s$-ratio
in all systems. For certain materials, e.g. helium, nitrogen and
water, the $\eta/s$-ratio has a minimum at the phase transition
\cite{K08}. Thus, there appeared belief that QCD nearly saturates
mentioned minimum near the critical point of the phase
transition~\cite{SZ04,CKML06}. These expectations agree with
estimates of the shear viscosity obtained within the lattice
QCD~\cite{KLP00,NS05,SN07,Mey07}. Actually, there is an extra
contribution of soft modes in the vicinity of the phase transition
critical point which may undergo a weak divergence, e.g. for the
first order liquid-gas phase transition ($H$ class of universality)
one expects the behavior
$\eta\propto |T-T_c|^{-\nu/19}\sim |T-T_c|^{-0.034}$, see \cite{Onuki}.
However, such a weak divergence can manifest itself only in a very
narrow vicinity of the critical point.

 The bulk viscosity is much less studied than the shear viscosity.
At high temperatures, when  coupling is weak and theory is nearly
conformal, the bulk viscosity is expected to be very
small~\cite{HK85,J95,JY95,ADM06,MS08}. In liquids the shear and
bulk viscosities are usually of the same order of magnitude.
However, in some cases the bulk viscosity can be significantly
higher than the shear viscosity. This is true in the presence of
soft slowly relaxing collective modes. For example, such behavior
occurs in the vicinity of the phase transition critical point or
at the crossover~\cite{MS08,ML37,LL06,PP06,KhT07,KKhT07}. In this
case divergence of the bulk viscosity is rather strong, $\zeta
\propto |T-T_c|^{-\nu z +\alpha}\sim |T-T_c|^{-1.8}$ (the critical
index $z$ is known from the $H$-model and $\nu,\alpha$ from $3d$
Ising model universality class)~\cite{Onuki}.  Lattice
calculations available for the gluon plasma are not in
disagreement with the expectation of an increase of the bulk
viscosity to the entropy density ratio,  $\zeta/s$, toward the QCD
phase transition critical point from above but error bars are
still very large~\cite{SN07,Mey08}. There are arguments, see
\cite{V93,SV09}, that the dynamics of the first-order phase
transition is controlled by finite values of the kinetic
coefficients.

In the modeling of the strongly interacting matter,  interactions
are often treated within the quasiparticle approximation.
Quasiparticle models~\cite{BKS05,Toneev} describe the lattice data
rather appropriately above the critical temperature $T_c$.
Relativistic mean-field based quasiparticle models were
successfully applied to describe the hadron phase
\cite{KTV07,KTV08}. In Refs.~\cite{ADM06,SR08} the shear and bulk
viscosities of the quark phase were  calculated within the quasiparticle
approach in the relaxation time approximation in the case where the
effective masses of the constituents depend on the temperature and
on the baryon density.

In this paper,  we extend the investigation of shear and bulk
viscosities  within the quasiparticle models in the relaxation
time approximation~\cite{HK85,ADM06,SR08}. In Sect. \ref{Rem}, we
describe the hadron phase ($T<T_c$) in terms of the quasiparticle
relativistic mean-field-based model with the scaling hadron masses
and couplings (SHMC)~\cite{KTV07,KTV08}. For the QGP phase in
Sect. \ref{HB} we use the "heavy quark bag" (HQB) model which
rather appropriately fits the lattice data. The equation of state
for the two-phase SHMC-HQB model  is constructed in Sect.
\ref{EoS}. In Sects. \ref{ViscH} and ~\ref{Quark} we evaluate the
shear and bulk viscosities for quasiparticle collisions in the
relaxation time approximation  and then  present results  of
numerical calculations first for the hadron and then  for
quark-gluon phases and compare them with previously obtained ones.
In Sect.~\ref{Res}, we estimate effects of finite mass-widths of
resonances on the viscosities and viscosity-to-entropy ratios. The
conclusion remarks are given in Sect. \ref{Concl}. Some details of
calculations are deferred to the Appendices A-C.

\section{The SHMC model setup}\label{Rem}
Consider  hadronic matter in thermal
equilibrium. Within our relativistic mean-field SHMC model we
present the Lagrangian density of the hadronic matter as a sum
of several terms:
 \be\label{math}  \mathcal{L}=\mathcal{L}_{\rm bar}+\mathcal{L}_
 {\rm MF}+\mathcal{L}_{\rm ex}~.
 \ee

 The  Lagrangian density of the baryon component interacting via
$\sigma,\omega$, $\rho$ mean fields  is as follows:
 \be
 \mathcal{L}_{\rm bar} &=& \sum_{b\in {\rm \{ bar \} }} \left[ i\bar \Psi_b\,
\Big(\prt_\mu +i\,g_{\om b} \,{\chi}_\om
 \ \om_\mu +ig_{\rho b}\chi_{\rho}\vec{\rho}_{\mu}\vec{t}_b\Big) \gamma^\mu\, \Psi_b -m_b^*\,
\bar\Psi_b\,\Psi_b
 \right] .
 \label{lagNn}
 \ee
  The considered baryon set is $\{b\}=N(938)$, $\Delta (1232)$,
$\Lambda (1116)$, $\Sigma (1193) $, $\Xi (1318)$, $\Sigma^*
(1385)$, $\Xi^* (1530)$, and $\Omega (1672)$, including
antiparticles. The used $\sigma$-field dependent effective masses
of baryons are~\cite{KTV07,KTV08,KV04}
\be \label{bar-m}
{m_b^*}/{m_b}=\Phi_b(\chi_\sigma  \sigma)= 1 -g_{\sigma b} \
\chi_{\sigma} \ \sigma /m_b \,, \; b\in\{b\}~.
 \ee
 In Eqs.
(\ref{lagNn}), (\ref{bar-m})  $g_{\sigma b}$, $g_{\om b}$,
$g_{\rho b}$ are coupling constants and $\chi_\sigma (\sigma)$,
$\chi_\om (\sigma)$, $\chi_{\rho}(\sigma)$ are coupling  scaling
functions.

The $\sigma$-, $\omega$-, $\rho$-meson contribution is
\begin{eqnarray} \mathcal{L}_{\rm
MF}&=&\frac{\prt^\mu \sigma \ \prt_\mu
\sigma}{2}-\frac{m_\sigma^{*2}\, \sigma^2}{2}-{U}(\sigma)
-\frac{\omega_{\mu\nu}\,\omega^{\mu\nu}}{4} +\frac{m_\om^{*2}\,
\om_\mu\om^\mu}{2}\nonumber\\
&-&\frac{\vec{\rho}_{\mu\nu}\,\vec{\rho}^{\mu\nu}}{4}
+\frac{m_\rho^{*2}\, \vec{\rho}_\mu\vec{\rho}^\mu}{2} ,
\\
 \omega_{\mu\nu}\,&=&\partial_\mu \om_\nu -\partial_\nu \om_\mu
 ~\,,\quad \vec{\rho}_{\mu\nu}\,=\partial_\mu \vec{\rho}_\nu
 -\partial_\nu \vec{\rho}_\mu~.
  \nonumber
\end{eqnarray}
The  mass terms of the mean fields are \be \label{bar-m1}
{m_m^*}/{m_m}&=&|\Phi_m (\chi_\sigma \sigma)|\,, \quad
\{m\}=\sigma,\om ,\rho\, . \ee

The dimensionless scaling functions $\Phi_b$ and $\Phi_m$, as well
as the coupling scaling functions $\chi_m$ depend on the scalar
field in the combination $\chi_\sigma(\sigma) \ \sigma$.
 Following~\cite{KV04} we assume approximate validity of the Brown-Rho
scaling ansatz in the simplest form
\be \label{Br-sc}\Phi =\Phi_N
=\Phi_\sigma =\Phi_\om =\Phi_\rho =
 1-f , \; \; \; f=g_{\sigma N} \ \chi_\sigma \ \sigma/m_N\,
 \ee
with $\chi_\sigma =\Phi_{\sigma}$.

 We keep a standard form for the  potential of the non-linear
self-interaction $U$ used in relativistic mean-field models,  now
expressed in terms of  $f$-variable:
\be
U&=&m_N^4 (\frac{b}{3}\,f^3 +\frac{c}{4}\,f^4 ) . \label{Unew} \ee

The third term in  the Lagrangian density (\ref{math}) includes
meson excitations
 \be
 \label{ex} \mathcal{L}_{\rm
ex}&=&\sum_{{\rm bos}\in {\{ \rm ex \} }}\mathcal{L}_{\rm bos}, \\ \ \
{\rm \{ ex\} } &=&\pi;K,\bar{K}; \eta (547);
\sigma',\omega',\rho';K^{*},\bar{K}^*(892),\eta'(958),\phi(1020).\nonumber
\nonumber
 \ee

The knowledge of the Lagrangian density (\ref{math}) defines
unambiguously the energy-momentum tensor $T_{\mu\nu}$ (Greek
indices  $\mu , \nu$ run $0,1,2,3$). The energy density $E$ and
pressure $P$ are given by the diagonal terms of the
energy-momentum tensor $ E=\left< T_{00}\right>, \
P=\frac{1}{3}\left<T_{ii}\right>$, Latin index $i=1,2,3$. The
pressure can be presented as a sum of the mean-field terms
and contributions of the baryons and  meson
excitations~\cite{KTV07,KTV08}:
 \be
\label{Efun} P[f,\om_0] &=& P_{\rm MF} [f, \om_0]
+\sum_{b\in\{{\rm bar}\}} P_b [f, \om_0] +
  \sum_{{\rm bos}\in\{{\rm ex}\}}P_{\rm
bos}[f,\om_0] ~.
 \ee
Below we consider isospin-symmetric nuclear matter ($N=Z$). Then
there are only $\sigma$ and $\om_0$ mean field solutions of
equations of motion. Therefore further we set the mean fields
$\vec{\om}=0$ and $\vec{\rho}_{\mu}=0$. The value of the mean
field $\om_0 (f)$ is found by minimization of the pressure and
then it is plugged back in the pressure functional. The latter
becomes a function only of $f$. The equilibrium value of $f$ can
be found by subsequent minimization of the resulting pressure in
this field. In a self-consistent treatment~\cite{KTV08}, equations
of motion for the mean
 fields render
 \be
  \frac{\prt}{\prt \om_0 }\,P [f,\om_0 ]=0\,~, \quad
\,\frac{d}{d f}\,P[f,\om_0 (f)]=\frac{\partial}{\partial f}\,
 P[f,\om_0 (f)]=0~.
  \label{extreme1}
   \ee
 Since the boson excitation term
 depends on the mean fields, its minimization produces extra
terms in the equations of motion for the mean fields.  This
self-consistency of the scheme allows us to be sure of
thermodynamic consistency of the model.

It is convenient to  introduce
 renormalized constants
 \be
C_m=\frac{m_N \ g_{mN}}{m_m}  \label{Cm}
 \ee
and, instead of $\chi_m$, consider other functions
 \be
\eta_{m}(f)={\Phi_m^2(f)}/{{\chi}_m^2(f)}\,\, \mbox{  for } \{m\}=\om ,\rho~,
\, \, \, \, \eta_\sigma (f)=1. \label{eta}
 \ee

In terms of this new notation the contribution of  mean fields
to the pressure (\ref{Efun}) is~:
 \be \label{Esig}
 P_{\rm MF} [f,\om_0]&=&-
\frac{m_N^4\,f^2}{2\, C_\sigma^2}\, \eta_{\sigma}(f) -{U}(f)+
\frac{m_N^2\eta_\om(f)}{2\,C_\om^2}\left[g_{\om
N}\,{\chi}_\om\,\om_0 \right]^2  .
 \ee
The density  of particle species $a$ is given by
\begin{eqnarray}
n_a (T,\mu_a ) = \frac{ \nu_a}{2\pi^2}\int_0^{\infty} d|\vec{p}|\ |\vec{p}|^2 \
F^{\rm eq}_a(k,T,\mu_a)~. \label{eqt1}
\end{eqnarray}
The corresponding  contribution
 to the pressure in (\ref{Efun}) is as follows:
 \be\label{Eb}
 P_a [f, \om_0]&=&
 \frac{\nu_a}{3}\intop_0^{\infty}\frac{\rmd |\vec{p}_a| \
|\vec{p}_a|^4}{2\pi^2}\,\frac{F_a^{\rm eq} }{E_a}~. \quad
 \ee
Here $a=(b\in\{\rm bar \}, {\rm {bos}} \in\{\rm ex\})$,
 "$b$" is the baryon and antibaryon, "${\rm bos}$" is the  boson and
 antiboson excitation as described above; $\nu_a$ is the degeneracy factor;
 \be\label{distr}
F_a^{\rm eq} =\left[e^{(E_a -\mu^*_a )/T}\pm 1\right]^{-1}
 \ee
 is the quasiparticle  occupation for  fermions ($+$) and bosons ($-$)
 with the quasiparticle energy
 \be\label{Eba}
E_a =\sqrt{(m_a^{\rm part*\,})^2+\vec{p}_a^{\,\,2}}~,
 \ee
effective mass $m_a^{\rm part*\,}$ and the gauge-shifted values of
chemical potentials
 \be \label{Tmu}
 \mu_{b}^* &=&t_b^{\rm bar}   \mu_{{\rm bar}}+t^{\rm str}_b  \
 \mu_{{\rm str}} -t_{b}^{\rm vec} \ g_{\om b} \ \chi_{\om} \ \om_0 = \mu_b -t_{b}^{\rm vec} \ g_{\om b}^* \om_0,
 \nonumber \\ \mu_{\rm bos}^* &=&t^{\rm str}_{\rm bos} \
 \mu_{{\rm str}}-t_{\rm bos}^{\rm vec} \ g_{\om ,\rm bos}^{*} \ \om_0
 =\mu_{\rm bos} -t_{\rm bos}^{\rm vec}
 \ g_{\om ,\rm bos}^{*}\ \om_0 .
 \ee
 Baryon (antibaryon) quantum numbers $t_b^{\rm bar}$, $t^{\rm str}_b$,
$t_{b}^{\rm vec}$ are the baryon charge, strangeness and
 vector charge, $t_{\rm bos}^{\rm vec}$ is the vector charge
 of the boson   (antiboson) excitation; $g_{\om ,\rm bos}^{*}$
are effective coupling constants.
We use the same values of the effective coupling constants and
masses  as in \cite{KTV08}.  One should specially note that as in
Ref. \cite{KTV08} here and below the coupling constants $g_{\sigma
b}$, except for nucleons, are additionally suppressed by factor
$1/10$ compared to  values which are usually used, see Fig. 3 in
\cite{KTV08}. This allows us to fit the lattice results for the
pressure of the quark phase up to $T\sim (220\div 230)$~MeV with
our SHMC model.

Summing up over  all relevant hadron species we obtain the
total baryon charge and strangeness in the hadron phase
\begin{eqnarray}
\label{eqt1b} n_{bar}(T,\mu_{\rm bar},\mu_{\rm str}) &=& \sum_{a
\in \{ h \}} b_a \ n_a(T,\mu_a)~, \\ n_{\rm str}(T,\mu_{\rm
bar},\mu_{\rm str}) &=&  \sum_{a \in \{ h\} } s_a \ n_a(T,\mu_a ),
\label{eqt1c}
\end{eqnarray}
 where the sum is taken over all hadrons $\{ h\}=\{ b
\}+\{{\rm bos} \}$.

The pressure of  boson excitations\footnote{We pay attention to a
misprint in Eq. (33)  of \cite{KTV08} for the kaon contribution.
In denominator enter $\om_K [\om_0=R_0=0]$.  } is the pressure of
the ideal Bose-gas of quasiparticles in the mean fields with
effective masses for $\sigma$-, $\omega$-, $\rho$- field
excitations and $K$ mesons, see (\ref{Eb}) -- (\ref{Tmu}).
For other particles we use bare masses.

In order to get $P_{\sigma}^{\rm part}$, we should expand total
pressure $P [\sigma ,\omega_0 (\sigma )]$  in $\sigma^{'} =\sigma
-\sigma^{\rm cl}$. The term linear in $ \sigma^{'}$ does not
contribute due to the subsequent requirement of the pressure
minimum in $\sigma^{\rm cl}$, $dP/df=0$. The quadratic term produces
 effective $\sigma^{'}$- particle mass squared,
 \be\label{Pspa1}
 (m_{\sigma}^{\rm part*})^2 =-\frac{{d^2 P_{\rm MF} [\sigma ,\omega_0
 (\sigma )]}}
 {{d \sigma^2}} =-\frac{{d^2 P_{\rm MF} [f ,\omega_0 (f )]}}
 {{d f^2}}\left(\frac{d f}{d \sigma}\right)^2 .
 \ee
 Keeping only quadratic terms in all
thermodynamical quantities in fluctuating fields we disregard the
boson excitation  and the baryon contributions in (\ref{Pspa1}).
Within our approximation  the effective masses of $\om$- and
$\rho$- excitations prove to be the same as those following from the
mean-field mass terms
 \be
 \label{omr} m_\om^{\rm part
  *} =m_\om |\Phi_\om (f)|, \quad m_\rho^{\rm part
  *} =m_\rho |\Phi_\om (f)|~.
\ee

We compare the results of our SHMC model with those for the ideal
gas (IG) of free particles. In the IG approximation the
density of the particle of the species $a$ with the spin-isospin
degeneracy factor $\nu_a$ is
\begin{eqnarray}
n_a^{\rm IG}(T,\mu_a ) = \frac{ \nu_a}{2\pi^2}\int_0^{\infty} dk\
k^2 \ F^{\rm IG}_a(k,T,\mu_a)~, \label{eqt1_IG}
\end{eqnarray}
where
\begin{eqnarray}
 F^{\rm IG}_a(k,T,\mu_a) = \left[ \
 \mbox{exp} \left( \frac{\sqrt{k^2 + m_a^2}- \mu_a}{T} \right)
 \pm 1 \right]^{-1}
\label{eqt2}
\end{eqnarray}
is the momentum distribution for fermions (upper sign $+$) and bosons
(lower sign $-$) in the IG limit. Here the chemical potential
\begin{equation}
\mu_a = b_a \ \mu_{\rm bar}+s_a \ \mu_{\rm str} \label{eqt1a}
\end{equation}
is related to the baryon $\mu_{\rm bar}$ and strange $\mu_{\rm
str}$ chemical potentials which control the conservation of the
baryon and strangeness charges, respectively; $b_a$, $s_a$ are the
baryon/antibaryon charge and strangeness for the given species.

\begin{figure}[thb]
\includegraphics[width=130mm,clip]{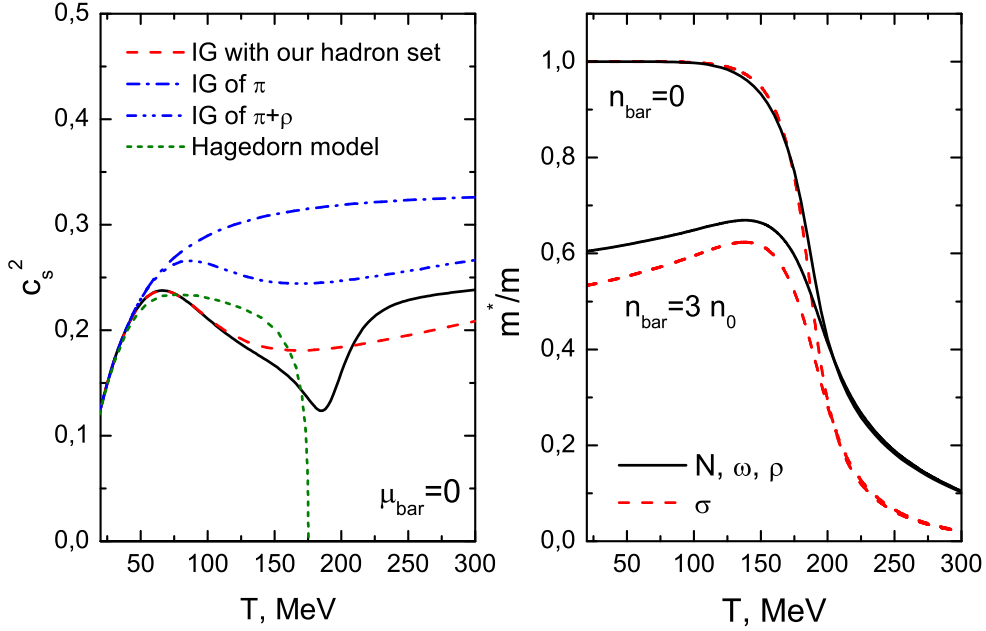}
\caption{ Left panel: The sound  speed  squared  in hadron matter
as a function of the temperature at zero baryon chemical
potential. Solid line -- calculation within the SHMC model.
Long-dashed line -- calculation for the  ideal gas of particles
performed with the same hadron set. Dash-dotted and
dash-double-dotted lines are results for the purely pion ideal gas
and ideal pion-rho gas,
 respectively. Calculation results within the Hagedorn gas
model~\cite{Hag65,CCMS09} are plotted by the short-dashed line.
Right panel: Temperature dependence of effective  masses for
nucleon, $\omega$- and $\rho$-excitations (solid line) and the
$\sigma$-meson excitation (dashed line) within the SHMC model for
two values of the baryon density.
 }
 \label{s-vel_h}
\end{figure}
In Fig.~\ref{s-vel_h} (left panel) we show the square of the speed
of  sound
 $c_s^2=dP/d\varepsilon$ at $\mu_{\rm bar} =0$ for the SHMC model
(solid line) and compare the result with that for the IG model
with different species number: with the same hadron set as in the
SHMC model (long-dashed line), for a $\pi +\rho$ mixture
(dash-double-dotted) and for a purely pion system (dash-dotted).
Short-dashed line shows $c_s^2$ for the Hagedorn model ({\em i.e.}
an IG of hadron resonances whose mass spectrum is assumed to have
the Hagedorn form $\rho (m)=m^{-a} \ \exp (m/T_{\rm H})$ where
$a=4$ and $T_H$ is interpreted as an upper bound of the hadron
temperature~\cite{Hag65,CCMS09}).

 As is seen, for the pure pion IG  the $c_s^2$
monotonously increases with increase of the temperature
approaching the ultrarelativistic  limit $c_s^2=1/3$ at high
temperatures. For the pion-rho meson mixture, the
$c_s^2$ exhibits a shallow minimum at $T\sim 170$ MeV. The minimum
(in the same temperature region)  is getting more pronounced for
multi-component systems (see long-dashed curve). At $T\lsim$ 50
MeV the pion contribution is a dominant one, thereby  all curves
coincide\footnote{Note that within the SHMC model pions are
treated as an ideal particle gas}. The curves for the SHMC model
and  IG  calculated with the same hadron set coincide for $T\lsim$
100 MeV. At $T>$ 50 MeV  heavier mesons start to contribute that
slows down  the growth of the pressure and then results in
decrease of $c_s^2$, contrary  to the case of the one-component
pion gas. At still higher temperatures heavier species start to
contribute in the case of multi-component systems resulting in  a
minimum in $c_s^2(T)$ for IG (see also \cite{CCMS09})  and SHMC
models. In the temperature  range $50$ MeV $<T<150$ MeV the
decrease rate of $c_s^2$ is slower  in the Hagedorn model than in
the SHMC and IG models. Only for $T\approx T_H >150$ MeV a
Hagedorn regime is reached, in which more and more energy goes
into forming massive resonances and therefore, finally $c_s^2$
drops to zero.
 As mentioned above, the use of the suppressed coupling
constants $g_{\sigma b}$, except for nucleons, guarantees that
 at $T_c < T<(220\div 230)$ MeV   the SHMC model results  for the
pressure are in agreement with the lattice results ~\cite{KTV08}.
Within the SHMC model with this modification, $c_s^2$ gets a deep
and rather narrow minimum at  a critical point $T\approx
T_c\simeq$ 180 MeV  \footnote{ The minimum of the speed of sound
(at $T= T_c \simeq$ 180 MeV) can be associated with a kind of
phase transition, e.g. with the hadron-QGP crossover, as it
follows from the detailed analysis of the lattice data, see
\cite{karsch06}.} caused by a sharp decrease of the in-medium
hadron masses at these temperatures (see the right panel in
Fig.~\ref{s-vel_h}, where effective masses of the nucleon,
$\omega$-, $\rho$- and $\sigma$-excitations are presented). Note
that in the $m$-truncated  Hagedorn-gas model, $c_s^2$ remains
finite at $T=T_H\approx T_c$: For $m<M = (1.5\div 2.5)$ GeV the
results~\cite{CCMS09} become very close to those for the IG model
presented in Fig.~\ref{s-vel_h}. The specific behavior  $c_s^2\to
0$ arises in the Hagedorn model only for $m\to \infty$.

\section{ The HQB model}\label{HB}

For the QGP phase of the IG of the massive quarks, antiquarks and
gluons  $\{q\}=q,\bar{q},g$,  see \cite{Toneev}  we define the density of
conserving charges similarly  to Eqs. (\ref{eqt1b}), (\ref{eqt1c}):
\begin{eqnarray}
\label{eqt11b}
 n_{\rm bar}^{\rm HQB}(T,\mu_{\rm bar},\mu_{\rm str})
&=& \sum_{a \in \{ q\} } b_a \ n_a^{\rm IG}(T,\mu_a )~, \\ n_{\rm
str}^{\rm HQB}(T,\mu_{\rm bar},\mu_{\rm str}) &=&  \sum_{a \in \{
q\} } s_a \ n_a^{\rm IG}(T,\mu_a)~. \label{eqt11c}
\end{eqnarray}

With nonperturbative effects associated with the deconfinement
transition included into a constant vacuum pressure $B$, the total
energy density and the pressure become
\be
 \varepsilon^{\rm HQB}(T,\mu_{\rm bar},\mu_{\rm str}) &=&
\sum_{a \in \{ q\}} \varepsilon_a^{\rm IG}(T,\mu_a) + B~,
\label{eqt10} \\
 P^{\rm HQB}(T,\mu_{\rm bar},\mu_{\rm str}) &=&  \sum_{a \in \{ q\}}
 P_a^{\rm IG}(T,\mu_a) - B~,
\label{eqt8}
 \ee
 where for gluons  ($a=g$) $\nu_g=16$. The gluon
chemical potential $\mu_g=0$; thereby, gluonic energy density
$\varepsilon$   and  pressure  $P$ depend only on temperature.

Effective masses of quarks and gluons are generally assumed to be
temperature and density independent and are treated here as free
parameters. From the fit
of thermodynamic quantities to available lattice data for the
$N_f=2+1$ system~\cite{karsch06,Cheng08,Sch08} we get quark masses
$m_u=m_d=100 $\, MeV, $m_s=450$ \, MeV and $m_g=600$\, MeV at
$B=(215 \ \mbox{MeV})^4$.

\section{Equation of state in the two-phase model}\label{EoS}

 Following a common strategy of the two-phase
model~\cite{Cleym86}, we determine here the deconfinement phase
transition by matching the EoS of the SHMC model and the HQB model
for quarks and gluons.

The phase equilibrium between the QGP and the hadronic phase at
the first order phase transition is determined by the Gibbs
conditions for thermal ($T^{\rm HQB}=T^{\rm SHMC}$), mechanical
($P^{\rm HQB}=P^{\rm SHMC}$) and chemical ($\mu_{\rm bar}^{\rm
HQB} =\mu_{\rm bar}^{\rm SHMC}, \ \mu_{\rm str}^{\rm HQB}
=\mu_{\rm str}^{\rm SHMC}$) equilibrium. The chemical equilibrium
between different components is automatically satisfied owing to
representation (\ref{eqt1a}) for the chemical potential. At given
temperature $T$ and baryon chemical potential $\mu_{\rm bar}$ the
strangeness chemical potential $\mu_{\rm str}$ is determined by
fixing the net strangeness of the  system to  zero. So, in the
Gibbs mixed phase at the fixed total baryon charge density $n_{\rm
bar}$ the following set of equations is to be solved~:
\begin{eqnarray}
&&P^{\rm SHMC}(T,\mu_{\rm bar},\mu_{\rm str}) =
P^{\rm HQB}(T,\mu_{\rm bar},\mu_{\rm str})~,  \nonumber \\
&&n_{\rm bar}(T,\mu_{\rm bar},\mu_{\rm str}) = \alpha \ n_{\rm
bar}^{\rm HQB}(T,\mu_{\rm bar},\mu_{\rm str})
 + (1- \alpha ) \ n_{\rm bar}^{\rm SHMC}(T,\mu_{\rm bar},\mu_{\rm str})~, \nonumber\\
&&0 = \alpha \  n_{\rm str}^{\rm HQB}(T,\mu_{\rm bar},\mu_{\rm
str}) + (1-\alpha ) \ n_{\rm str}^{\rm SHMC}(T,\mu_{\rm
bar},\mu_{\rm str})~, \label{eqt12a} 
\end{eqnarray}
where $\alpha = V^{\rm HQB} /V$ is a volume fraction occupied by
plasma, {\it i.e.} the coexistence phase is not a homogeneous one.
The boundaries of this mixed phase are found by putting $\alpha =
1$ ({\it the hadron phase boundary}) and $\alpha = 0$ ({\it the
plasma phase boundary}).   Eqs. (\ref{eqt12a}) should be solved
at every phase point $(T,n_{\rm
bar})$ of the coexistence region. It results in that the
quark-hadron boundary of a system conserving baryon charge and
strangeness becomes a critical surface
(not a line in the $T-\mu_{\rm bar}$ plane as for a
single charge conservation, but some stretched-out
area~\cite{TNFN04}.)

\begin{figure}[thb]
 \hspace*{2mm}
\includegraphics[width=65mm,clip]{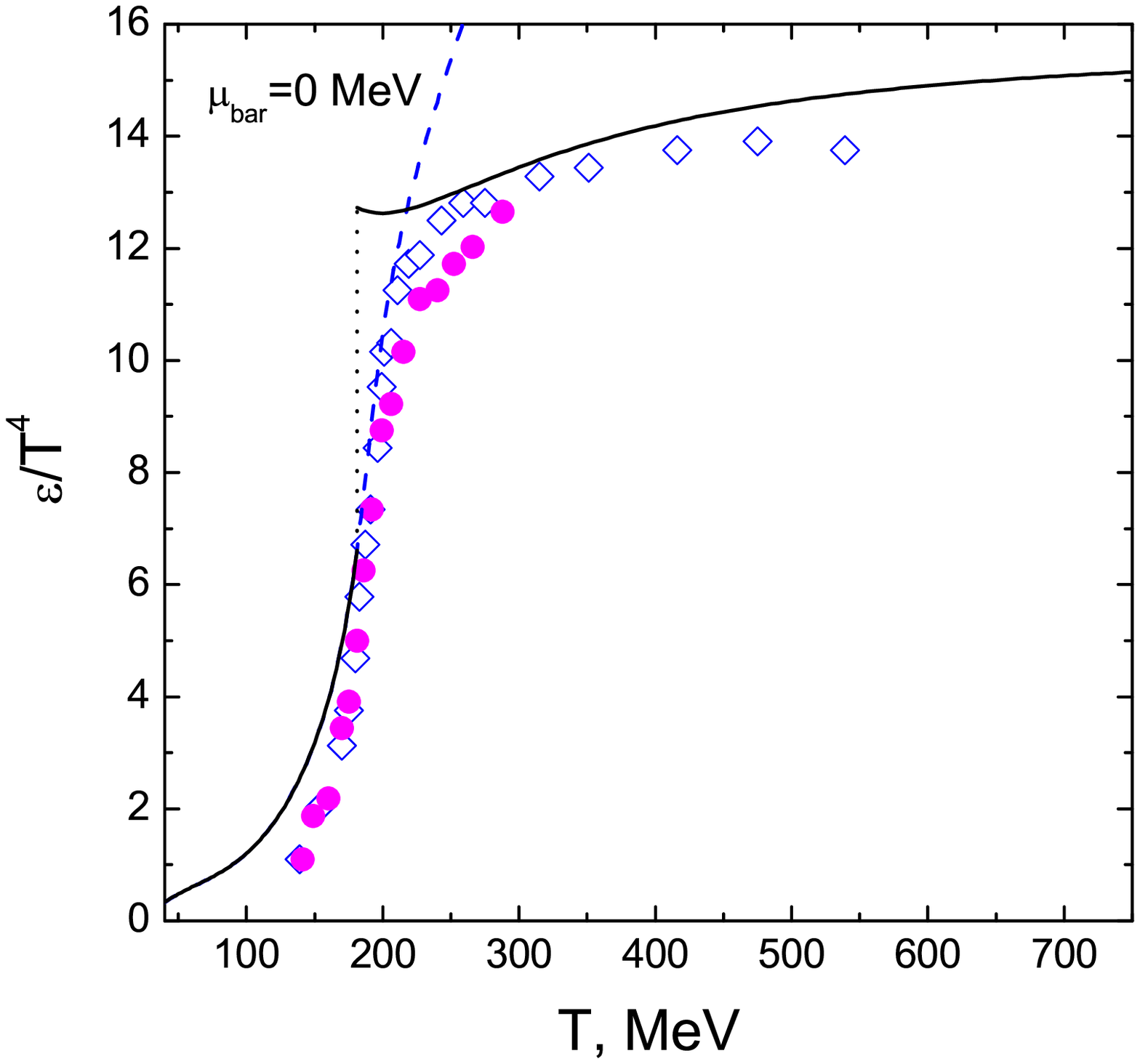}
\includegraphics[width=65mm,clip]{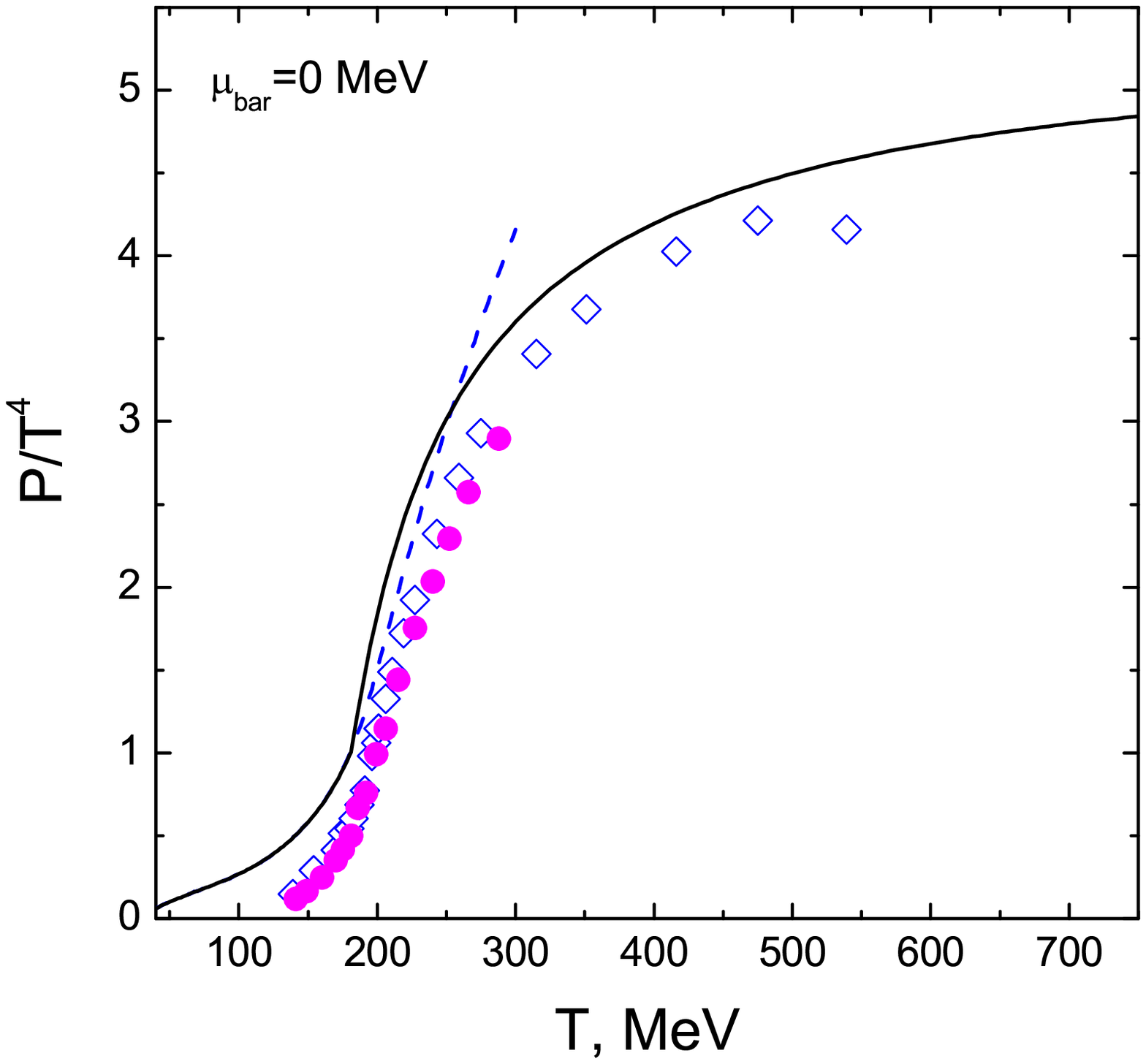}
\caption{ Reduced energy density and pressure at $\mu_{\rm bar}=0$
obtained  in the two-phase SHMC-HQB model. The energy density jump
at $T_c\approx 180$ MeV  marked by the dotted line corresponds to the
mixed phase. The hadronic SHMC results  above $T_c$ are
plotted by  dashed lines. QCD lattice results with the time extent $N_t=$8
for the p4 and asqtad actions are plotted by empty diamonds and filled
circles, respectively~\cite{Cheng09}.
}
 \label{epT4}
\end{figure}

 Predictions of the hadronic SHMC model  with suppressed
couplings agree with the recent lattice data for the reduced
energy and the pressure  up to about $T\simeq 220\div 230$ MeV
(see Fig.~\ref{epT4}, left and right).
 As follows from Fig.~\ref{epT4},
at higher temperatures quark-gluon degrees of freedom should be
taken into account. Thereby, further we use the SHMC-HQB mixed
phase model.
In the  $\mu_{\rm bar}=0$  case the energy density of the SHMC-HQB
mixed phase model suffers a jump at the  temperature
$T=T_c \approx 180$ MeV (the left panel in Fig.~\ref{epT4}).
This value of the critical temperature is consistent with that
calculated on the lattice  at $n_{\rm bar}=0$~\cite{Cheng08,Cheng09}.
However the lattice data demonstrate
 a sharp crossover transition at this point. We found a jump in
the latent heat about $\Delta \varepsilon \sim 0.85 $~GeV/fm$^3$
($\Delta\varepsilon/T^4 \sim 6$). The presence of the jump in
$\varepsilon$ at $T_c$ indicates the first order phase transition.
From the right panel of Fig.~\ref{epT4} we see that the pressure
undergoes a continuous monotonous increase in temperature in agreement with
the construction of the mixed phase.

\begin{figure}[thb]
 \bc
\includegraphics[width=65mm,clip]{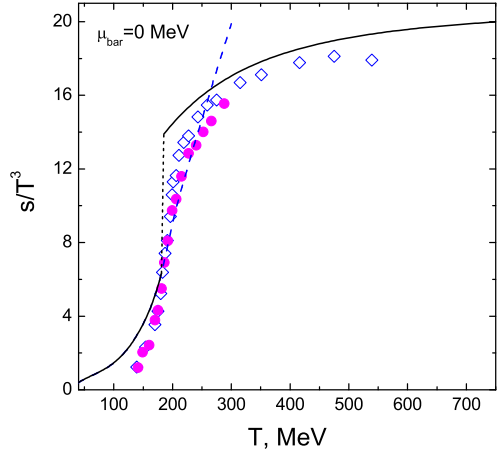}
\caption{Reduced entropy density  at $\mu_{\rm bar}=0$ obtained
in the two-phase SHMC-HQB model. The entropy density jump at
$T_c\approx 180$ MeV marked by the dotted line corresponds to the
mixed phase. Notation is the same as in Fig. \ref{epT4}.}
 \label{sT3}\ec
\end{figure}

In Fig. \ref{sT3} we demonstrate  the reduced entropy density as a
function of temperature. The lattice data using two types of the
action are plotted by points for $N_t=$8. The results of the two-phase
model  and purely hadronic SHMC model are shown by solid and
dashed (prolonged for $T>T_c$) lines, respectively.  The SHMC
model with suppressed couplings continues to be in close agreement
with the lattice data till $T\sim$230 MeV.
The HQB model agrees reasonably with the lattice
data~\cite{Cheng09} for all temperatures above $T_c$ except a
narrow vicinity of the critical point where in the lattice QCD we
deal with the crossover at $\mu_{{\rm bar}}=$0.
\begin{figure}[thb]
 \hspace*{2mm}
\includegraphics[width=130mm,clip]{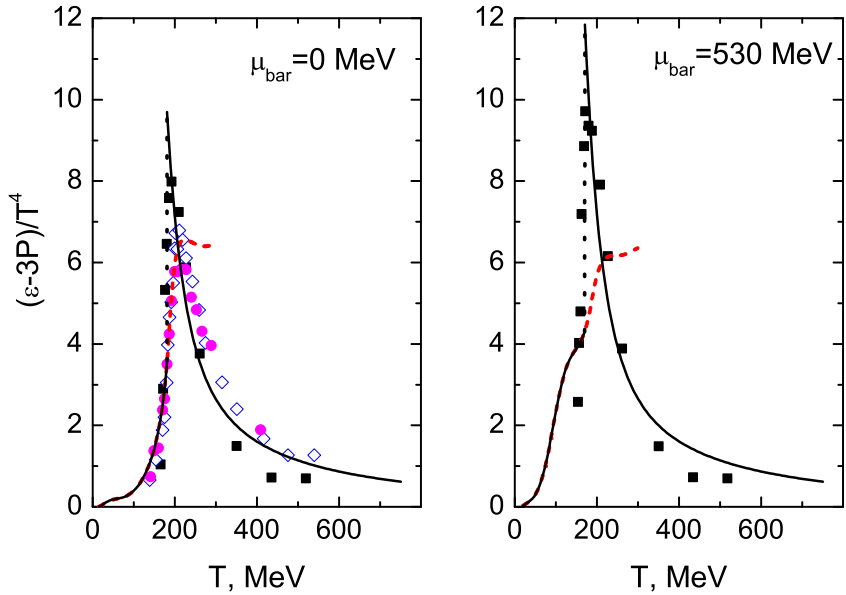}
\caption{ Trace anomaly at $\mu_{\rm bar}=0$ and $\mu_{\rm
bar}=530$ MeV.  The jump is marked by the dotted line. The
hadronic SHMC results are plotted  above $T_c$ by  dashed lines.
Lattice QCD data are from \cite{fod02} (filled squares), and
from~\cite{Cheng09}  (circles and diamonds, notation is as in Fig.
\ref{epT4}).
 }
 \label{e3p}
\end{figure}

Reasonable agreement between the SHMC-HQB model and lattice data
is observed  not only for the reduced energy, pressure  and
entropy but also for the trace anomaly $(\varepsilon-3P)/T^4$,
for both zero and finite values of the baryon chemical potentials,
as demonstrated in the left and right panels of  Fig.~\ref{e3p}.
As is seen, only at large temperatures
$(\varepsilon-3P)/T^4\rightarrow 0$, i.e. the system becomes a
conformal one. The conformal invariance in QCD is essentially
broken at $T\lsim 2T_c$. Therefore, we may expect here differences
in predictions of  conformal theories for different quantities,
e.g. for $\eta/s$,  at $T\lsim 2T_c$.  We stress that  the
conformal regime is not reproduced by the purely hadronic SHMC
model,  valid roughly for $T\lsim  230$ MeV.

\begin{figure}[thb]
 \hspace*{2mm}
\includegraphics[width=130mm,clip]{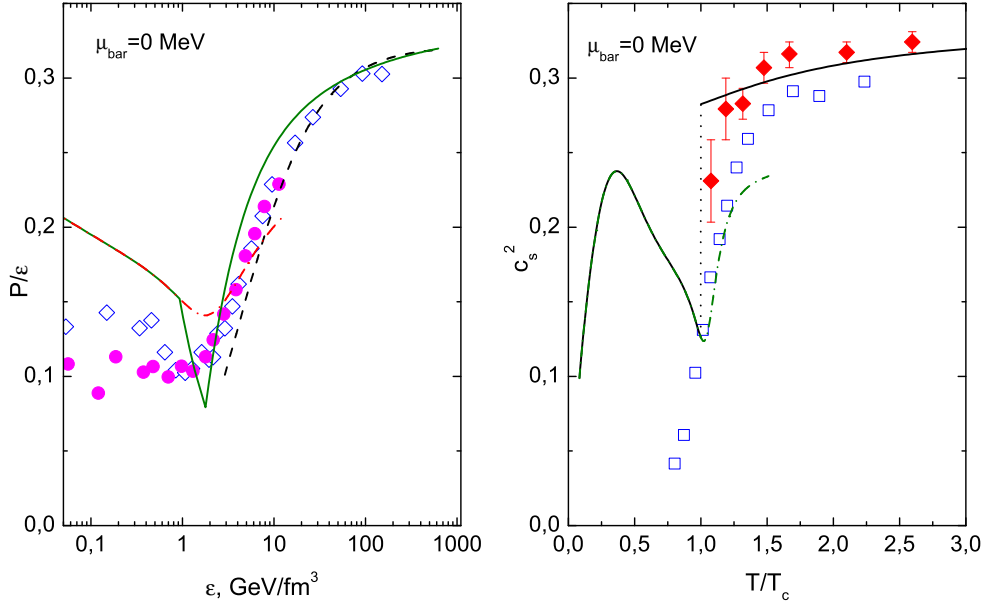}
\caption{ The $P/\varepsilon$ ratio demonstrating the presence of
the softest point and the speed of sound squared in the baryonless
matter are presented. Solid lines in both panels are results of
our two-phase model calculations.  Dash-dotted lines
continuing the solid lines at $T>T_c$ demonstrate calculation
results of the SHMC model. Left: empty diamonds and filled
circles are the $N_f=$2+1 lattice data for the $P/\varepsilon$
ratio calculated with the p4 and asqtad actions,
respectively~\cite{Cheng09}. The dashed line is the high-$T$
approximation (\ref{appr1}) of lattice data with improved
staggered fermions and almost physical masses for
$N_f=2+1$~\cite{Cheng08}. Right: Diamonds and squares are lattice
QCD data for $N_f=$2 with Wilson fermions~\cite{karsch06} and
$N_f=$2+1 with staggered fermions~\cite{AFKS05}, respectively.
  }
 \label{vs2}
\end{figure}

Values of  the ratio $P/\varepsilon$ and the speed of
sound $c_s$ essentially determine the evolution of a system, and
as it will be shown below, the last quantity is an important
ingredient entering into the expression for the bulk viscosity
coefficient.

In Fig.~\ref{vs2} the energy density dependence of the
$P/\varepsilon$ ratio for the given two-phase  SHMC-HQB model EoS
is presented alongside with the temperature dependence of $c_s^2$.
Both quantities ($P/\varepsilon$ and $c_s^2$)  have close physical
meaning and coincide in the ultrarelativistic limit
$c_s^2=P/\varepsilon =1/3$. This limiting trend is observed in
Fig.~\ref{vs2} (left and right panels) at high temperatures and
energies. In the high energy density region the SHMC-HQB results
agree reasonably well with the lattice data. Though the cited
above lattice results were obtained in different discretization
schemes for 2- and (2+1) flavor QCD using different quark masses
and lattice spacing, thermodynamics in the high temperature phase
is not strongly influenced by discretization errors and is rather
insensitive to changes of the quark mass. Nevertheless, one should
be cautious treating  these lattice QCD results for not too high
  temperatures, especially for $T<T_c$. Thereby, our point
here is to use the HQB model for $T>T_c$ that fits the lattice
results well for $T\gsim 2T_c$ and to apply our SHMC model
in the hadronic sector at $T<T_c$, instead of using the
corresponding lattice values for $T\lsim T_c$.  It is of interest
to note that the SHMC model reasonably reproduces both
$P/\varepsilon$ and $v_s^2$ at the temperature above $T_c$ till
about 230 MeV, in agreement with results presented in Figs.
\ref{epT4},\ref{sT3},\ref{e3p}.

Lattice data~\cite{Cheng08} on the $P/\varepsilon$  ratio obtained
 for the pion mass $m_\pi\approx$ 220 MeV are well
approximated in the high temperature/energy range
$1.3\lsim\varepsilon^{1/4} \lsim 6 \ $ GeV$/$fm$^3$ as
 \be
 \frac{P}{\varepsilon}=\frac{1}{3} \left(0.964-\frac{1.16}{1+
 0.26 \ \varepsilon \ \mbox{fm}^3/\mbox{GeV}} \right)~.
 \label{appr1}
\ee
 This approximating curve is plotted by the dashed line in
Fig.~\ref{vs2} (left panel).  We see that different lattice
data presented in this figure agree rather well with each other.
Since $c_s^2=dP/d\varepsilon$,  the approximation
(\ref{appr1}) can also be used to reproduce the speed of sound.
It is seen that
in contrast with a single component hadronic gas (cf. Fig.~\ref{s-vel_h}
and Fig.~\ref{vs2}), both $c_s^2$ and $P/\varepsilon$ exhibit a
minimum near the phase transition. In other words, the EoS in our
two-phase SHMC-HQB model is getting softer near $T_c$ and the
system evolves slower at the ``softest point'', i.e. at the
minimum of the $P/\varepsilon$ ratio. As was noted above, for hadronic
sector $T<T_c$ the lattice results predict much lower values of
the $P/\varepsilon$ ratio, as compared to our SHMC model.
 Being in
discrepancy with lattice results for the low-temperature hadronic
component,  results of our two-phase and SHMC models agree
reasonably well
 with other hadronic models, as it is demonstrated in
Fig.~\ref{s-vel_h}.

 The following remarks are in order. As  was found
within the standard RMF model~\cite{Theis83} the specific heat
exhibits a sign of critical behavior. The SHMC model with the
standard choice of couplings demonstrates a crossover at a
pseudocritical temperature $T\sim T_c$ for $\mu_{\rm bar}=0$ in
agreement with the lattice data. In this case we observe a sharp
peak in the heat capacity which is typical for the strong
crossover transition~\cite{KTV07}.   In our case when $g_{mb}$
couplings are suppressed, the peak is smoothed that reminds about
a weak crossover. Our two-phase SHMC-HQB model shows the first
order phase transition.  This can be treated as an advantage of
the SHMC model and a disadvantage of the two-phase model. However
the results of the SHCM model  begin to disagree with the lattice
data for $T>220-230$~MeV, whereas the two-phase model fits well
the lattice data in the high temperature region. Thus we can't use
the SHMC model for high temperatures without its significant
modification. We refuse of doing such a modification bearing in
mind that the high temperature regime should be treated in terms
of the quark-gluon degrees of freedom. The mentioned lack of our
two-phase model can be removed by a more careful treatment of the
near-critical region. On the one hand, in the framework of the model with
the first-order phase transition our consideration of the vicinity
of the critical point is obviously oversimplified, since the Gibbs
condition's treatment disregards  dynamical effects  in the
phase transition. The latter effects are important for the
description of the vicinity of the critical end point. These
dynamical effects are manifested  at temperature below the
critical end point in all models with the EoS  of the van der
Waals type, see \cite{SV09,Csernai}.  In particular, an
overcooled state to be possibly formed at the first order phase
transition may result in a mechanism of fast
hadronization~\cite{Mish}. Thus one way to better treat the
near-critical region is to incorporate non-trivial dynamical
effects, e.g. in the framework of the van der Waals model EoS. On
the other hand, one could smooth the thermodynamic quantities in
the near-critical region in such a way that the EoS would
correspond to the crossover for sufficiently small values of
$\mu_{\rm bar}$, see \cite{Brazil}. We will return to the
consideration of the critical effects in the subsequent work.

 \section{Shear and bulk viscosities in the hadron phase}
\label{ViscH}

\subsection{Collisional viscosity in the SHMC model. Derivation of equations}

Following Sasaki and Redlich \cite{SR08}, we derive expressions
for the shear and bulk viscosities  in the case when the
quasiparticle spectrum is given by the expression
$E(\vec{p})=\sqrt{\vec{p}^{\,2} +m^{*\,2}(T,\mu)}$. We perform a
similar derivation, but in the presence of mean fields $\sigma$
and $\om_0$. In the latter case one should additionally take into
account that quasiparticle distributions depend on the mean
fields.

We start with the expression for the energy-momentum tensor:
\begin{eqnarray}
T^{\mu\nu}=T^{\mu\nu}_{\rm MF}+\sum_{b\in \{\rm bar\}}T^{\mu\nu}_b
+ \sum_{{\rm bos}\in \{\rm ex\}}T^{\mu\nu}_{\rm bos},
\label{tmunuT}
\end{eqnarray}
where the mean-field contribution is as follows
\begin{eqnarray}
&&T^{\mu\nu}_{\rm MF}=
\left(\frac12\left[
{m^*_\sigma}^2 \
\sigma^2 -{m^*_\omega}^2 \ \omega_0^2\right]+U(\sigma)
\right)g^{\mu\nu},
\end{eqnarray}
with $m^*_\sigma$ and $m^*_\omega$  given by Eq.~(\ref{bar-m1}),
and where we dropped the terms quadratic in gradients.

The quasiparticle (fermion and boson excitation) contribution is
\begin{eqnarray}\label{Tmunua}
T^{\mu\nu}_a =\int d\Gamma\left\{ \frac{(p_a^\mu +X_a^\mu )
p_a^\nu}{E_a}F_a \right\}.
\end{eqnarray}
Here
\begin{eqnarray}
 p_a^\mu &=& (E_a (\vec{p}_a ,\vec{r}_a),\vec{ p}_a),\quad  d\Gamma =
 \nu_a \frac{d^3p_a}{(2\pi)^3},\\
 X_b^\mu &=& g_{\om b} \ \chi_{\om} \ \om^\mu ,\quad   X_{\rm ex}^\mu =
 g_{\om ,\rm ex}^* \ \om^\mu.\nonumber
\end{eqnarray}
 Note that the contribution to the energy-momentum tensor
$T_a^{\mu\nu}$ is not symmetric with respect to the interchange of
indices. The tensor can be symmetrized following a general rule,
e.g. see \cite{Weinberg}: One may add to $T^{\mu\nu}$ an extra
term $T^{'\mu\nu }$, such that $\partial_{\mu}T^{'\mu\nu }=0$ and
that $T^{\mu\nu}+T^{'\mu\nu}=T^{\nu\mu}+T^{'\nu\mu}$. Thus the
asymmetry of $T_a^{\mu\nu}$ does not influence on observables. In
addition, to calculate viscosities we need to deal only with
spatial components of the tensor, $T^{ik}$. Since $X^\mu =(X_0,
0,0,0)$ in the  mean field approximation used by us, up to second
gradients we have $T^{ik}_a \simeq T^{ki}_a$. Therefore we will
not perform this symmetrization procedure in the given paper.

 The  quasiparticle distribution function $F_b$ for baryon
components in the presence of mean fields fulfills the Boltzmann
kinetic equation \cite{SCFNW},
\begin{eqnarray}\label{Boltz1}
\left(p^{\mu}_b\partial_{\mu}-g_{\om
b}p_{\mu}\om^{\mu\nu}\frac{\pd }{\pd
{p}^{\nu}_{b}}+m_b^{*}\partial^{\nu}m_b^{*}\frac{\pd }{\pd
{p}^{\nu}_{b}}\right)\widetilde{F}_b=St \widetilde{F}_b ;
\end{eqnarray}
with $\widetilde{F}_b  ({p}_b ,x_b) =\delta (p_b^2 -m_b^{*\,2})
F_b^{\rm loc.eq.}(\vec{p}_b ,x_b) $.

   The local equilibrium boson or baryon distribution is given as follows:
 \be
 F^{\rm loc.eq.}_a (\vec{p}_a ,x_a )=\left[e^{(E_a -\vec{p}_a \vec{u}-
\mu_a + t_a^{\rm vec}X_a^0 )/T}\pm 1 \right]^{-1},\quad X_a^0 =
g_{\om a} \ \chi_{\om} \ \om_0 ,
 \label{leqdf}
 \ee
where we suppressed $\vec{u}^{\,2}$ terms for $|\vec{u}|\ll
1$.\footnote{ For $\vec{u}\neq 0$ there appear mean field
solutions with $\vec{\om}\neq 0$. These however yield
$\vec{u}^{\,2}\ll 1$ terms.} Here the upper sign ($+$) is for
fermions and ($-$) is
 for bosons, and the vector particle charge is
 $ t_a^{\rm vec}=\pm 1$ or $0$; $g_{\om a}\neq 0$ only for $a\in \rm bar$
 and kaons in our model.
Considering only slightly inhomogeneous  solutions and using
$|\vec{u}|\ll 1$ we may drop the terms $\propto \vec{u}^2$ and
$\propto \vec{u}\nabla \om_0$ in the kinetic Eq. (\ref{Boltz1}).
Then kinetic equations for boson and baryon components acquire
ordinary quasiparticle form
\begin{eqnarray}\label{Boltz}
\frac{\pd F_a}{\pd t}+\frac{\pd E_a}{\pd \vec{p}_a}\frac{\pd
F_a}{\pd\vec{r}_a}-\frac{\pd E_a}{\pd \vec{r}_a}\frac{\pd
F_a}{\pd\vec{p}_a}=\frac{p^{\mu}_a}{E_a} \ \frac{\partial
F_a}{\partial x^{\mu}_a}=St F_a ,
\end{eqnarray}
 where  $p_a^\mu = (E_a (\vec{p}_a ,\vec{r}_a ,\sigma,\om ),\vec{ p}_a)$.
  We used  that
$\pd E_a /\pd \vec{p}_a=\vec{p}_a/ E_a$.
 Since  calculating the viscosity,  we need only terms with
velocity gradients, we further put $\pd E_a/\pd \vec{r}_a=(\pd
E_a/\pd \mu_a)\vec{\nabla}_a\mu_a +(\pd E_a/\pd T)\vec{\nabla}_a T
=0$.

 In the relaxation time approximation
 \be
 \label{st}
 St F_a =-\delta F_a /\tau_a , \quad\delta F_a = F_a -F_a^{\rm loc.eq.} .
 \ee
 Here $\tau_a$ denotes the relaxation time of the given
species. Generally, it depends on the quasiparticle momentum
$\vec{p}_a$.

 The averaged  partial relaxation time ${\tilde\tau}_a$ is related to
 the cross section   as
 \be
  {\tilde \tau}^{-1}_a (T,\mu )
=\sum_{a^{'}} n_{a^{'}} (T,\mu)\left<v_{aa^{'}}
\sigma^t_{aa^{'}}(v_{aa^{'}}) \right>,
 \label{tau}
 \ee
where $n_{a^{'}}$ is the density of $a^{'}$-species,
$\sigma^t_{aa^{'}}=\int d\cos \theta \ d\sigma(aa^{'}\to
aa^{'})/d\cos \theta \ (1-\cos \theta) $ is the transport cross
section, in general,  accounting for in-medium effects and
$v_{aa^{'}}$ is the relative velocity of two colliding particles
$a$ and $a^{'}$ in  case of binary collisions. Angular brackets
denote a quantum mechanical statistical average over an
equilibrated system. In reality, the cross sections entering the
collision integral and the corresponding relaxation time $\tau_a$
in (\ref{st}) may essentially depend on the particle momentum.
Thus, averaged values ${\tilde\tau}_a^{-1}$ given by Eq.
(\ref{tau}) yield only a rough estimate for the values
${\tau}^{-1}_a$ which we actually need for calculation of
viscosity coefficients, see below Eqs. (\ref{shear}) and
(\ref{bulk}).

 In the relaxation time approximation from
 Eqs. (\ref{Boltz}), (\ref{st}) we obtain
\be\label{deltaF}
 \delta F_a =-\frac{\tau_a}{E_a } \ p^{\mu}_a  \ \frac{\partial
 F_a^{\rm loc.eq.}}{\partial
x^{\mu}_a},
 \ee
 and then the nonequilibrium correction to the energy-momentum tensor
(\ref{tmunuT}) becomes:
 \be
 \label{varenmom} \delta T^{\mu\nu} &=&-\sum_a\int
d\Gamma\left\{ \tau_a\frac{(p_a^\mu +X_a^\mu )
p_a^\nu}{E_a^2} \ p_a^\kappa\pd_\kappa F_a \right\}_{\rm loc.eq.}\\
&+&\delta\sigma \ \frac{\partial T^{\mu\nu}}{\partial\sigma}_{\rm
loc.eq.} +\delta\om_0 \ \frac{\partial
T^{\mu\nu}}{\partial\om_0}_{\rm loc.eq.}~.\nonumber
 \ee
Considering small deviations from the local equilibrium, we may
keep in (\ref{varenmom}) only first-order derivative quasiparticle
terms $\propto \partial_{\mu}$.

The shear and bulk viscosities expressed through traceless
nondiagonal and diagonal parts of the nonequilibrium correction
to the energy-momentum tensor  are   as follows:
 \be
 \label{vis}
\delta T_{ij}&=&-\zeta \ \delta_{ij}{\vec{\nabla}}\cdot\vec{u}-\eta \ W_{ij},\\
 \nonumber
W_{kl}&=& \pd_ku_l+\pd_lu_k-\frac23\delta_{kl} \ \pd_iu^i
.
 \ee
  As before, here Latin indices run $1,2,3$.

To find the shear viscosity,  we  put $i\neq j$ in (\ref{vis})
 and use that in this case  the variation of the second and
third terms in (\ref{varenmom}) yields zero after integration over
angles. To find the bulk viscosity, we  substitute $i=j$ in
(\ref{vis}) and use that $T^{ii}_{\rm eq}=3P_{\rm eq}$.  As
follows from equations of motion (\ref{extreme1}), variation of
the second and third terms in (\ref{varenmom})   again yields
zero. We put $\vec{u}=0$ in final expressions but retain gradients
of the velocity.

Taking derivatives  $\partial F_a^{\rm loc.eq.}/\partial
x^{\mu}_a$ in Eq. (\ref{deltaF}) with the help of Eqs. (\ref{Tt})
-- (\ref{mstrt}) from Appendix A, we  find the variation of the
total energy-momentum tensor as the function of  derivatives of
the velocity \footnote{ Since we are not interested in
calculations of the heat conductivity, we again suppress the
$\vec{\nabla} T$, $\vec{\nabla} \mu$ terms and  also put
$\vec{u}=0$ keeping only the corresponding derivative terms.}
 \be
 \label{deltaTij}
\delta T^{ij}&=&\sum_{a}\int d\Gamma \ \frac{p^i_a
p^j_a}{TE_a}\,\tau_a \ F_a^{\rm eq}(1\mp  F_a^{\rm eq})\,q_a (\vec
p\,;T,\mu_{\rm bar} ,\mu_{\rm str})~
 \ee
 with the upper sign ($-$) in the blocking factor for fermions and  the lower
 one ($+$) for bosons in accordance with Eqs.~(\ref{eqt2}),(\ref{leqdf}),
 \be
 q_a (\vec p\,;T,\mu_{\rm
bar} ,\mu_{\rm str} )=\pd_ku_l \ \delta_{kl} \ Q_a
-\frac{p_kp_l}{2E_a} \ W_{kl},
 \ee
\begin{align}\label{Qa} &Q_a =
-\left\{\frac{\vec{p}_a^{\,2}}{3E_a}+\left(\frac{\pd P}{\pd n_{\rm
bar}}\right)_{\epsilon,n_{\rm str}}\left[\frac{\pd (E_a
+X^0_a)}{\pd\mu_{\rm bar}}-t_b^{\rm bar}
\right]\right.\nl&\left.+\left(\frac{\pd P}{\pd n_{\rm str}
}\right)_{\epsilon,n_{\rm bar}}\left[\frac{\pd (E_a
+X^0_a)}{\pd\mu_{\rm str}}-t_a^{\rm str}\right]-\left(\frac{\pd
P}{\pd \epsilon}\right)_{n_{\rm bar} ,n_{\rm str} }\right.\\
&\left.\times\left[E_a +X^0_a -T\frac{\pd (E_a +X^0_a )}{\pd
T}-\mu_{\rm bar}\frac{\pd (E_a +X^0_a)}{\pd \mu_{\rm bar}
}-\mu_{\rm str} \frac{\pd (E_a +X^0_a)}{\pd \mu_{\rm str}
}\right]\right\}.\nonumber
\end{align}

Finally, we obtain the shear viscosity
\begin{align}\label{shear}
\eta&=\frac{1 }{15T}\sum_{a}\int d\Gamma\,\tau_a
\frac{\vec{p}^{\,4}_a}{E^2_a}\,\left[ F^{\rm eq}_a \; (1\mp F^{\rm
eq}_a )\right].
\end{align}
Correspondingly, the bulk viscosity  is
\begin{align}
\label{bulk}
\zeta&=-\frac{1}{3T}\sum_{a} \int\mathrm d\Gamma\,\tau_a \;
\frac{\vec{p}^{\,2}_a}{E_a} \; F^{\rm eq}_a \;
\left(1\mp F^{\rm eq}_a\right)Q_a .
\end{align}

To get a final expression for the latter quantity, one should use the
energy conservation obtained from Eq. (\ref{deltaTij}) by
demanding $\delta T^{00}=0$,
\be
\label{encons}
\sum_a\int &d&\Gamma\, \left[\tau_a \; (E_a+X_a^0) \; F_a^{\rm eq} \;
(1\mp F_a^{\rm eq}) \; q_a \right. \nonumber \\  &+& \left. \bar\tau_a \;
(E_a-X_a^0)\; \bar F_a^{\rm eq} \; (1\mp \bar F_a^{\rm eq})
 \; q_{\bar a}\right]=0~,
\ee
where the second term corresponds to antiparticles with the relaxation
time $\bar{\tau}_a$.
After substitution of Eq. (\ref{encons}) in (\ref{bulk}), the
bulk viscosity can be presented as
\begin{align}
\zeta&=-\frac{1}{3T}\sum_a\int\mathrm
d\Gamma\,\left(-\frac{{m_a^*}^2}{E_a}-X_a^0\right)\tau_a \,
F_a^{\rm eq}\,\left(1\mp F_a^{\rm eq}\right) \;
Q_a\nl&-\frac{1}{3T}\sum_a\int\mathrm
d\Gamma\,\bar\tau_a\left(-\frac{{m_a^*}^2}{E_a}+X_a^0\right)\bar
F_a^{\rm eq}\,\left(1\mp\bar F_a^{\rm eq}\right) \, Q_{\bar
a},
\label{bulkT}
\end{align}
where $Q_a$ is defined by Eq. (\ref{Qa}).

For $X_a^0 =0$, $\mu_{\rm str}=0$ our results coincide with those
obtained in \cite{SR08}:
\be
\label{etaC} \eta_a&=&\frac{1 }{15T}\int d\Gamma\,\tau_a
\frac{\vec{p}^{\,4}_a}{E^2_a}\,\left[ F^{\rm eq}_a \; (1\mp F^{\rm
eq}_a )\right],\\
\label{zetaC} \zeta_a&=& -\frac{1}{3T}\int\mathrm
d\Gamma\,\tau_a\frac{{m_a^*}^2}{E_a} \, F_a^{\rm
eq}\,\left(1\mp F_a^{\rm eq}\right)\;\times\nl&\times&\left\{
\frac{\vec{p}_a^{\,2}}{3E_a}+\left(\frac{\pd P}{\pd n_{\rm
bar}}\right)_{\epsilon,n_{\rm str}}\left[\frac{\pd E}{\pd\mu_{\rm
bar}}-t_b^{\rm bar} \right]\right.\\&&\left.-\left(\frac{\pd
P}{\pd \epsilon}\right)_{n_{\rm bar} ,n_{\rm str}
}\left[E_a-T\frac{\pd E_a}{\pd T}-\mu_{\rm bar} \frac{\pd E_a}{\pd
\mu_{\rm bar} }\right]\right\}
\nonumber
 \ee
where the quasiparticle mass $m_a^*$ can be temperature/density
dependent. Making substitution of the on-shell mass instead of
$m_a^*$ and $F^{\rm eq}_a=F^{\rm IG}_a$ in Eqs. (\ref{etaC}),
(\ref{zetaC}) we arrive at expressions for the gas  of
non-interacting particles.
For the gas of quasiparticles with  values $m^*$ depending only on
the temperature with the help of Eq. (\ref{encons}),  Eq.
(\ref{zetaC}) can be rewritten in a simpler  form \cite{G85}
\be
\label{zetaCT} \zeta&=& \frac{1}{T}\int\mathrm d\Gamma\,\tau\,
F^{\rm eq}\,\left(1\mp F^{\rm eq}\right)\;\left\{ \left(
\frac13-c_s^2\right)E-\frac{{m^*}^2}{3E}\right\}^2 .
 \ee
  In a particular case of the one component gas of material particles
 in the high temperature limit $T\gg m$ for $\tau ={\tilde\tau} =const$
viscosity coefficients take the simplest  form \cite{G85}:
 \be
 \eta \simeq \frac{4 {\tilde\tau}}{15}\,e(T),\quad \quad  \quad \quad \zeta \simeq 0,
\label{eta0}
 \ee
 where $e(T)=\frac{1}{30}\pi^2 T^4$ is the known
 black-body radiation result for a massless Bose fluid.

 \subsection{Collisional viscosity in hadronic baryonless
matter}

 In the relaxation time
approximation both shear and bulk viscosities for a component
"$a$" \ depend on its relaxation (collisional) time $\tau_a$,
which should be parameterized or calculated independently.
Therefore to diminish this uncertainty it is legitimate at first
to find the reduced kinetic coefficients (per unit relaxation
time, assuming $\tau =const$, i.e. $\tau ={\tilde\tau}$).

\begin{figure}[h]
 \hspace*{2mm}
\includegraphics[width=120mm,clip]{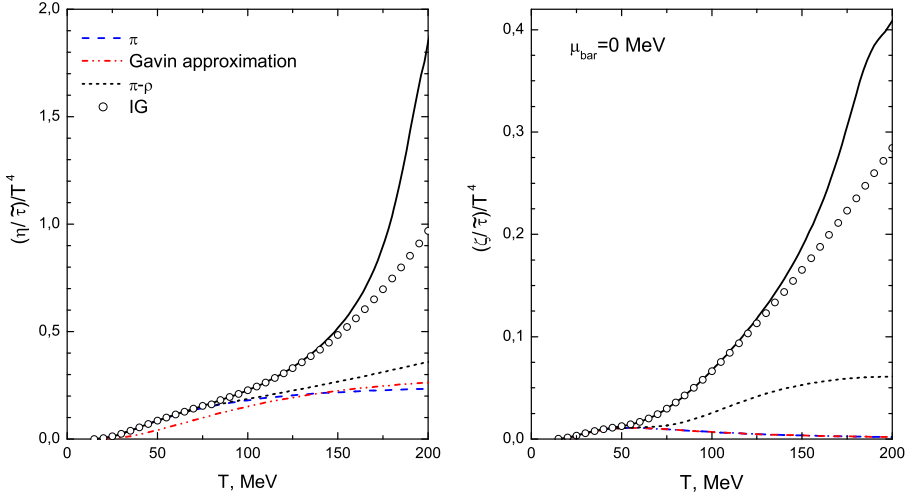}
\caption{  The reduced $T^4$ scaled shear (left panel) and bulk
(right panel) viscosities per unit relaxation time calculated
within the SHMC model for a system with $\mu_{\rm bar}=$0 (solid
lines). The results are compared with those for the  massive pion
gas (dashed lines), $\pi-\rho$ mixture (short dashed line)
calculations and with some interpolation  from  massive to massless pion gas
(the Gavin approximation~\cite{G85}, dash-double-dotted
line) as well as for the IG model with the same large set of
species as in the SHMC model (open dots).
 }
 \label{Gav}
\end{figure}

In Fig.~\ref{Gav} we demonstrate  results of various calculations
for the reduced shear (left panel) and  bulk  (right panel)
viscosities at $\mu_{\rm bar}=0$ additionally scaled by the $1/T^4$
factor. As we see from the figure, the reduced scaled shear viscosity
of the massive pion gas (dashed line) becomes approximately constant
 for $T\gsim$ 100 MeV. Naturally,  this result is close to
that obtained in the Gavin approximation~\cite{G85} where a
numerical interpolation between the $m/T\ll 1$  and $m/T\gg 1$
cases was used (dash-double-dotted line in Fig.~\ref{Gav}).   The
$T^4$ scaling is violated for the $\pi-\rho$ gas in the
temperature interval under consideration because the $\rho$ mass
is not negligible even at $T\sim $ 200 MeV. For $\zeta$ the
approximate $1/T^4$ scaling property holds for the massive
pion-rho gas at $T\gsim$ 150 MeV. Note that $\zeta=$0 for the  gas
of free massless pions since $c_s^2=1/3$ in this case. For the
massive pion gas $\zeta/T^4$ gets maximum for $T\simeq 60$ MeV and
then begins to decrease for $T>60$ MeV reaching zero at large $T$
similar to the case of massless gas.
The reduced shear and bulk  viscosities of a multicomponent system
calculated in  our SHMC model (solid lines) and in the IG model
with the same hadron set (open dots) do not fulfill the $T^4$
scaling law. These models include  a large set of hadrons.  Due to
that with the temperature increase  the reduced shear and bulk
viscosities    become significantly higher
 than those for the pion gas and the
pion-rho gas models. An additional increase of the reduced
viscosity within the SHMC model originates from  significant mass
decrease at temperatures near the critical temperature. The bulk
viscosity of a single-component pion system drops to zero both at
low and high temperatures and in the whole temperature interval
$\zeta<<\eta$, that is frequently used as an argument for
neglecting the bulk viscosity effects. However, the statement does
not hold anymore for  mixture of many species. For example, at
$T\sim$150 MeV the $\eta/\zeta$ ratio is only about 3 in the case of
the IG and SHMC models.  Thus the bulk viscosity effects can play
a role in the description of  the hadronic stage at high collision
energies, like at RHIC. Moreover, the bulk viscosity can be
responsible  for such important effect as flow anisotropy.

 For further evaluation of the absolute values of the transport
coefficients $\eta_a$ and $\zeta_a$ we need to  estimate the
relaxation time of hadronic species.  For that we use Eq.
(\ref{tau}).  We apply free cross sections  in the case of
the IG based model, similar to procedure performed in
Ref.~\cite{PPVW93}. For the SHMC model,  the in-medium
modification of cross sections is incorporated by a shift of a
``pole'' of the collision energy by the mass difference
$m_a-m_a^*$ according to prescription of Ref.~\cite{BC08}. Due to
a lack of microscopic calculations this is the only modification,
which we do here.

\begin{figure}[thb]
 \hspace*{20mm}
\includegraphics[width=90mm,clip]{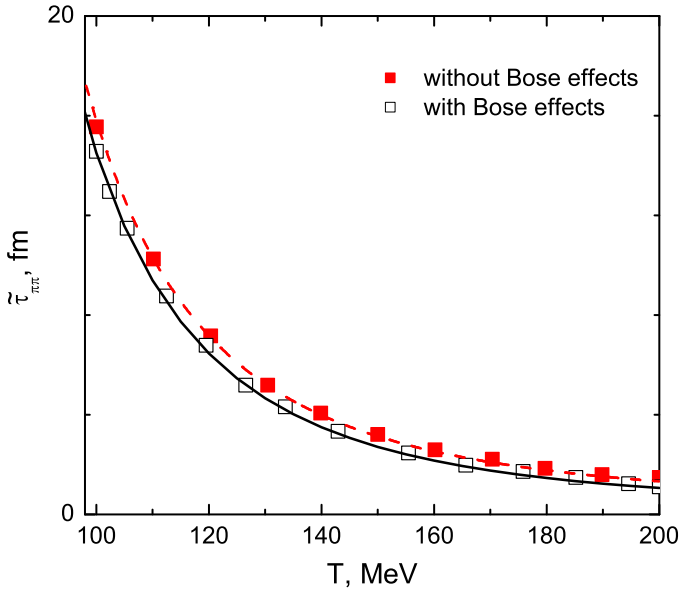}
\caption{The pion relaxation time  for the $\pi\pi$  system
 as a function of temperature.  Solid and dashed
lines are our calculation results for quantum Bose and classical
Boltzmann statistics, respectively. The corresponding points (empty
and black squares)  are taken  from the
review-article of Prakash et al.~\cite{PPVW93}.
 }
 \label{tau_pi}
\end{figure}

The relaxation time for the  $\pi\pi$ collisions in the purely
pion gas is plotted in Fig.~\ref{tau_pi}. The same cross sections
as in Ref.~\cite{PPVW93} are taken since pions within our SHMC
model are assumed to have  free dispersion law, as well as in Ref.
\cite{PPVW93}. Therefore our results (solid and dash  curves)
 coincide with those in Ref.~\cite{PPVW93}  (empty and black
squares). As is seen, there is  only a small influence of quantum
statistics on the pion relaxation time.

\begin{figure}[thb]
 \hspace*{10mm}
\includegraphics[width=110mm,clip]{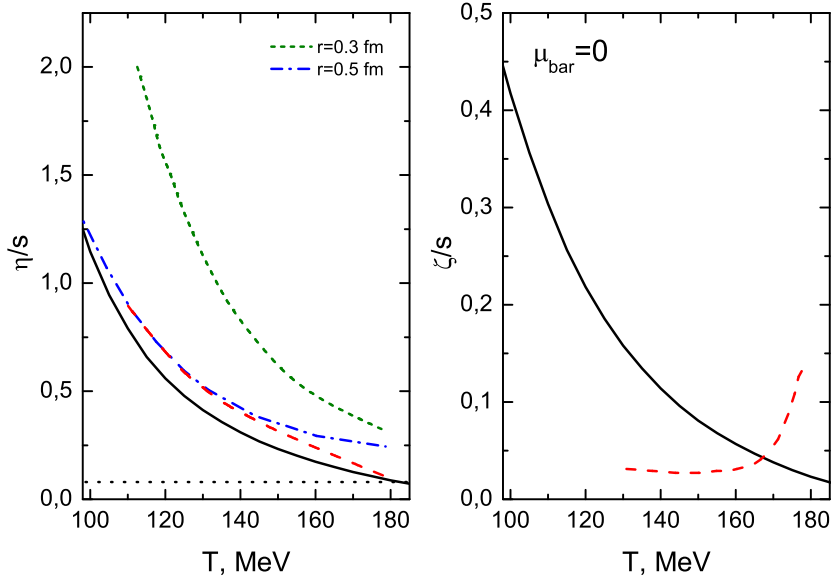}
\caption{Specific shear (left panel) and bulk (right panel)
viscosities ($\eta/s$ and $\zeta/s$) of hadrons as a function of
$T$ for $\mu_{\rm bar}=$0. Our SHMC model results are shown by
solid lines. The results of the excluded volume hadron gas
model~\cite{GHM08} are shown for two values of hard-core radii
${\rm r}$ (short dashed and dash-dotted lines). The results for the
resonance gas model including Hagedorn states \cite{NHNG08} are
shown by the long-dashed lines. The dotted horizontal line
corresponds to the lower AdS/CFT bound, $\eta/s=1/4\pi$~\cite{KSS03}.
 }
 \label{sh_b_T}
\end{figure}

The temperature dependence of the ratio of transport coefficients
to the entropy density is analyzed in Fig.~\ref{sh_b_T} for  the
$\mu_{\rm bar}=$0 system. Here we make emphasis on the high-$T$
behavior to clarify the question how hadronic models describe
approaching the lower bound of the $\eta/s=1/4\pi$ ratio predicted
by strongly coupled theories~\cite{PSS01,BLS05,KSS03}. As is seen
from Fig.~\ref{sh_b_T} (left panel), the SHMC model (solid lines)
as well as the excluded volume one (short dashed and dash-dotted
lines)~\cite{GHM08} predict a monotonous decrease of the specific
shear viscosity $\eta/s$. However, only in the SHMC model the
lower bound $\eta/s \sim 1/4\pi$ is reached at the critical
temperature $T_c\approx 180$ MeV.   The
 hadron model~\cite{NHNG08} describes all the known particles
and resonances with masses $m_i<2$ GeV  in terms of an excluded
volume model with $r=0.5$ fm and additionally includes an
exponentially increasing number of Hagedorn states for $m_i>2$
GeV. It is natural that for
 $T\lsim $140 MeV the resonance gas model with the
Hagedorn states reproduces the behavior of $\eta/s$ for the hadron
excluded volume model for $r=0.5$ fm. At $T\sim T_c$  results for the
shear viscosity calculated in the line of classical non-relativistic
approximation~\cite{NHNG08} are consistent
with the lower bound of the $\eta/s$
ratio. Note that our SHMC relativistic mean-field model does not
need extra assumptions on  the particular Hagedorn states to
describe the dip in $c_s^2$ and $\eta/s \sim 1/4\pi$ near the
critical temperature.

 The specific bulk viscosity $\zeta/s$ (see Fig.~\ref{sh_b_T}, right panel)
behaves quite differently in  our SHMC and the Hagedorn
 hadronic models since completely different assumptions were
 used for calculation of the bulk  viscosity.
Namely, our results are based on the quasiparticle collisional
relaxation time approximation while in ~\cite{NHNG08}, following
the QCD sum rules in~\cite{KhT07,KKhT07}, the bulk viscosity
$\zeta$ was related with the trace anomaly $<\theta>_T\equiv
<T^\mu_\mu>_T=\varepsilon-3P$:
 \be
\zeta(T)=\frac{1}{9\omega_0(T)}\left[T^5\frac{\partial}{\partial
T} \frac{<\theta >_T-<\theta >_0}{T^4}+16|\varepsilon_0|\right],
\label{trace}
 \ee
where $|\varepsilon_0|=<\theta >_0/4$ in vacuum. To derive Eq.
(\ref{trace}), a particular ansatz was used for the spectral
function at zero   momentum. However, recently this ansatz
has been criticized in~\cite{MS08,Mey08,HKP08,RS09}.  Moreover,
 Eq. (\ref{trace}) obviously is not applicable for temperatures
far from $T_c$.

Viscosities in the interacting hadronic phase were calculated in
Ref.~\cite{CW07} for $T<T_c$ when   the dominant configuration of
QCD with two flavors of massless quarks is the interacting gas of
chiral pions. In the large $N_c$ limit for massless chiral pions
it was obtained that
 \be \left(\frac{\zeta}{\eta}\right)_{\pi}\sim
(\frac{1}{3}-\frac{P}{\varepsilon}) \ (\frac{1}{3}-c_s^2)~.
 \label{csf}
 \ee
 As was first shown by Gavin \cite{G85} in pure massless ideal
 pion gas $\zeta =0$. However if pions are coupled with themselves
 or with
the matter of other species $\varepsilon\neq  3P$ and
$\zeta_{\pi}\neq 0$. Eq. (\ref{csf}) is similar to a simplified
expression $\zeta /\eta\sim 15(\frac{1}{3}-c_s^2)^2$ which is
obtained for a photon gas coupled to  hot matter~\cite{MLK05} and
is also parameterically correct for perturbative QCD~\cite{ADM06}.
This is because $2\rightarrow 2$ scattering is the dominant
process in both $\eta$ and $\zeta$ calculations. It is however not
the case, {\it e.g.} in the $\lambda \phi^4$ model, where $\eta$
is dominated by
 $2\leftrightarrow 4$ processes while
$\zeta$  by $2\rightarrow 2$.  For a strongly coupled $N=2^{*}$
gauge theory using AdS/CFT \cite{BLS05} the scaling is $\zeta/\eta
\propto (\frac{1}{3}-c_s^2)^2$.

\begin{figure}[thb]
 \hspace*{20mm}
\includegraphics[width=80mm,clip]{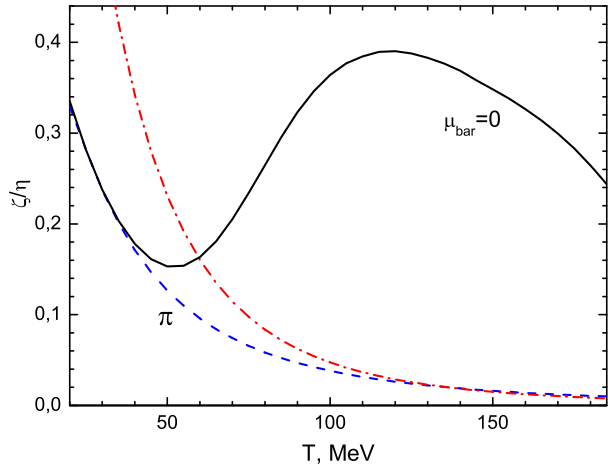}
\caption{ The ratio of the bulk-to-shear viscosity for  baryonless
hadronic matter  calculated within the SHMC model (solid line) and
for the  massive ideal pion gas (dashed line). For comparison
$T$-dependence of the ratio calculated following  Eq.~(\ref{csf})
(i.e. as for chiral pions, but with thermodynamical quantities as
for the massive ideal pion gas and normalized to the value for the
massive pion gas at large $T$), is shown by the dash-dotted line.
 }
 \label{vis_ratio}
\end{figure}

  In Fig.~\ref{vis_ratio} we show the ratio $\zeta/\eta$
calculated for baryon-less matter in SHMC model (solid line) and
those computed for  the massive pion gas with a free dispersion
law (dashed). For comparison the $\zeta/\eta$-ratio calculated
according to Eq.~(\ref{csf}) (i.e., as for chiral pions)  but with
thermodynamic quantities the same as for the massive pion gas is
shown by  the dash-dotted line.  At $T\gsim m_{\pi}$ pions become
almost chiral ones and the $T$ dependence of the pion gas indeed
can be described by Eq. (\ref{csf}). This coincidence of two
curves at large $T$ allowed us to estimate the value of the
numerical pre-factor  in Eq. (\ref{csf}).
 For $\mu_{\rm bar} =0$ the
$\zeta/\eta$ ratio in our SHMC model coincides at $T< 50$ MeV with
that for the free massive pion gas. It demonstrates that the main
contribution to viscosities is given here by the pion mode. In the
purely pion case the ratio $\zeta/\eta$ monotonously decreases
with increase of the temperature, whereas within our SHMC model it
gets a minimum at $T\sim 50$ MeV and then a maximum for $T\sim
120$ MeV.

 \subsection{Collisional viscosity in baryon enriched hadronic
matter}

\begin{figure}[thb]
 \hspace*{2mm}
\includegraphics[width=65mm,clip]{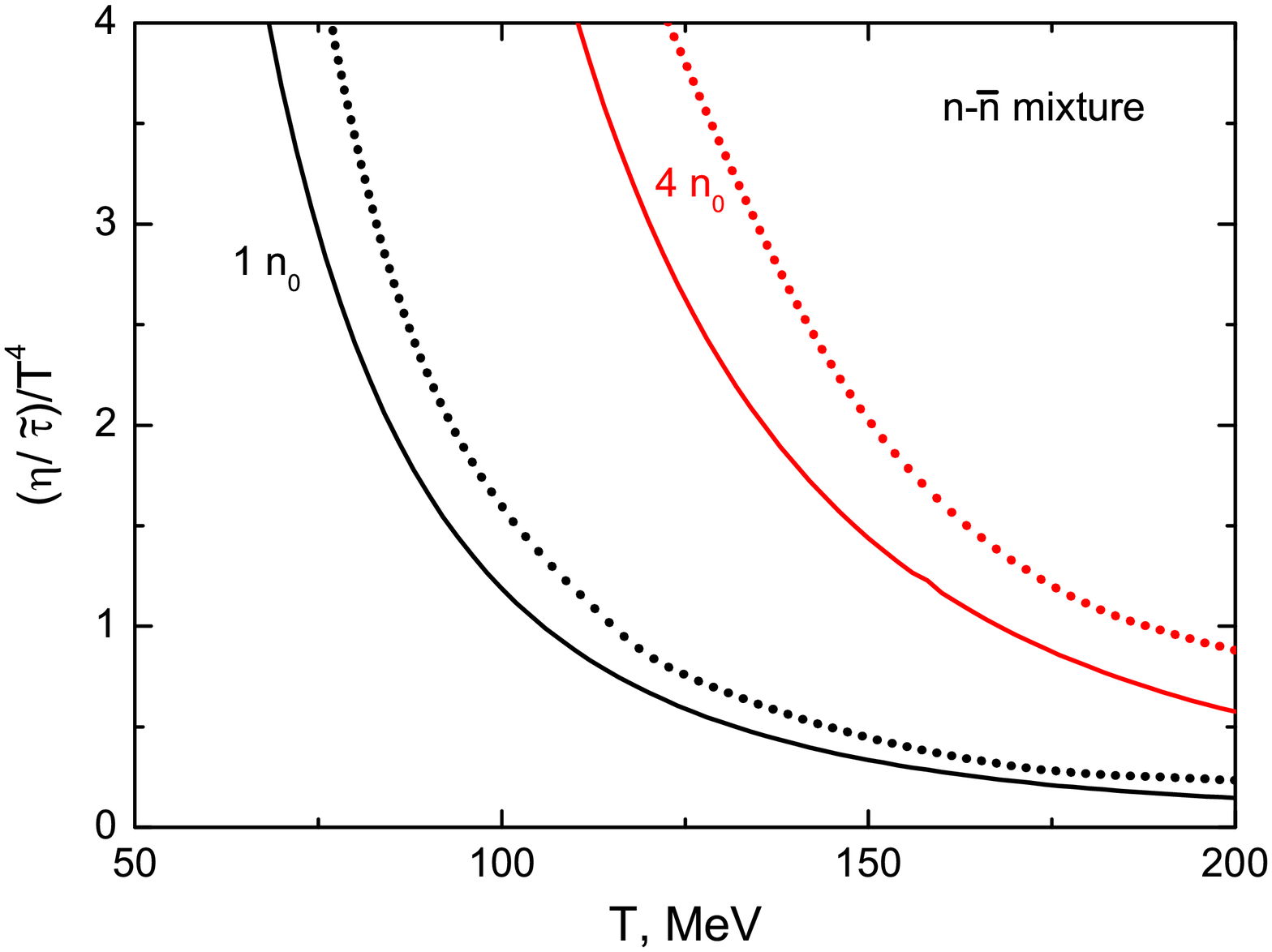}
\includegraphics[width=65mm,clip]{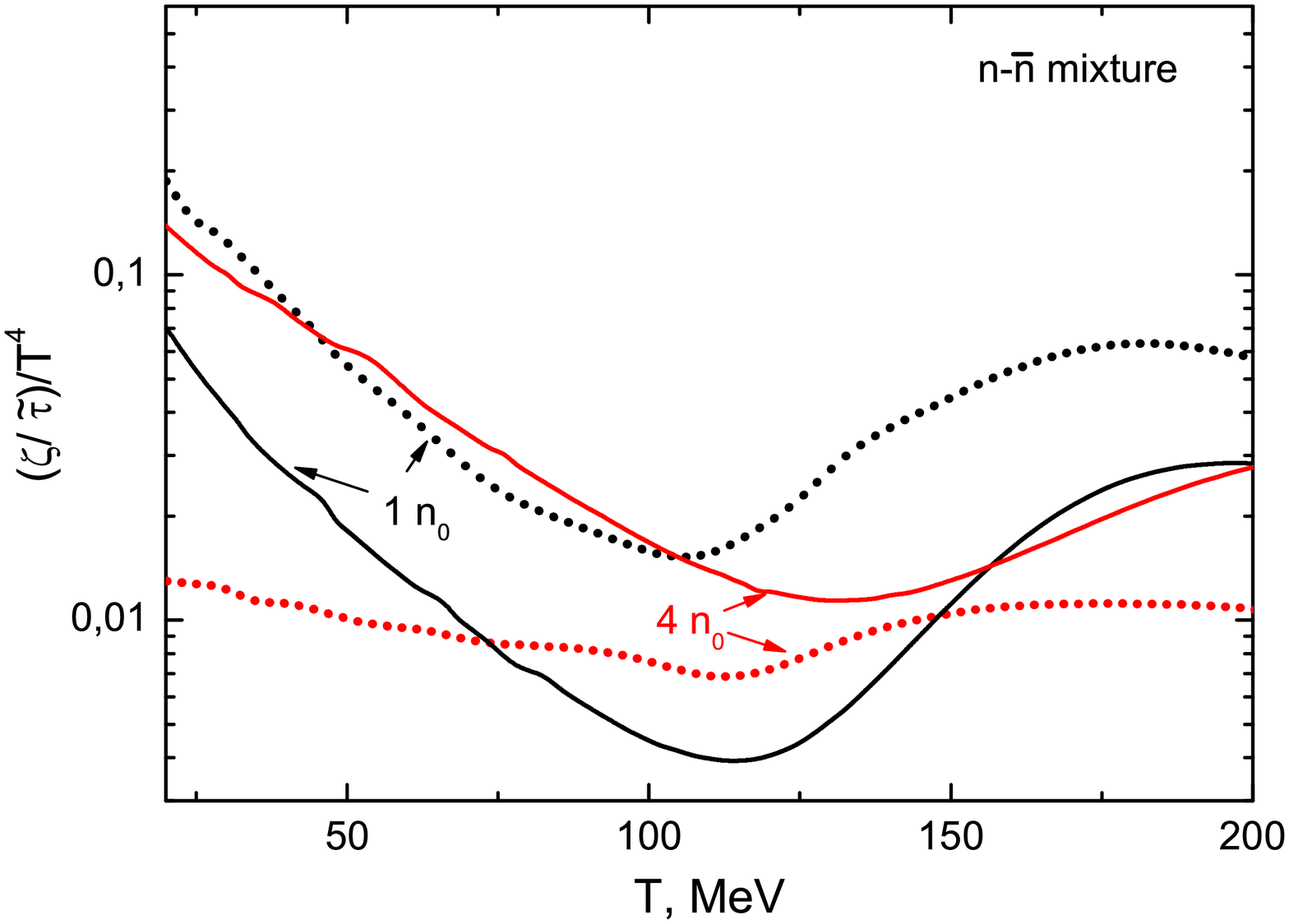}
\caption{The temperature dependence of the reduced  $T^4$-scaled
shear (left panel) and bulk (right panel) viscosities per unit
relaxation time  calculated for a one-component
 pure nucleon-antinucleon system at $n_{\rm bar}=n_0$
and $4n_0$. Solid lines are the SHMC model calculations. The
standard Walecka model results~\cite{HM93} are plotted by dotted
lines.
 }
 \label{eta_N}
\end{figure}

Similarly to the $\mu_{\rm bar}=$0 case, to avoid  influence of
uncertainties in the value of the relaxation time we study at
first  reduced kinetic coefficients. Transport coefficients for a
 nucleon-antinucleon mixture
 at finite baryon density were calculated earlier by
Hakim and Mornas within the standard Walecka model~\cite{HM93}.
Comparison between their (dotted lines) and our (solid lines) SHMC
model results is given in Fig.\ref{eta_N} for two values of the
density. Since Ref. \cite{HM93} presented results for one
component (neutron) matter, we performed calculations in  the SHMC
model also for neutron-antineutron matter. Because the $\rho$ mean
field is absent in the standard Walecka model, we suppressed the
$\rho$-term  also  in the SHMC model calculation. We see that the
reduced $T^4$-scaled shear viscosity behaves similarly in both
models but absolute values differ due to different effective
masses.  Only at high temperatures and for $n_{\rm bar} \simeq
n_0$ ($n_0$ is the nuclear saturation density)   the values of
reduced shear viscosities become close in both models. Differences
in the values of the reduced  bulk viscosity calculated in two
models are larger than for the reduced shear viscosity because
$\zeta$ depends on the specific behavior of thermodynamical
quantities (cf. Eqs. (\ref{etaC}) and (\ref{zetaC})). In
particular, it concerns  the high-density region (see the curves
for $n_{\rm bar}= 4n_0$), when the mass in the standard Walecka
model is getting very small.

\begin{figure}[thb]
 \hspace*{2mm}
\includegraphics[width=65mm,clip]{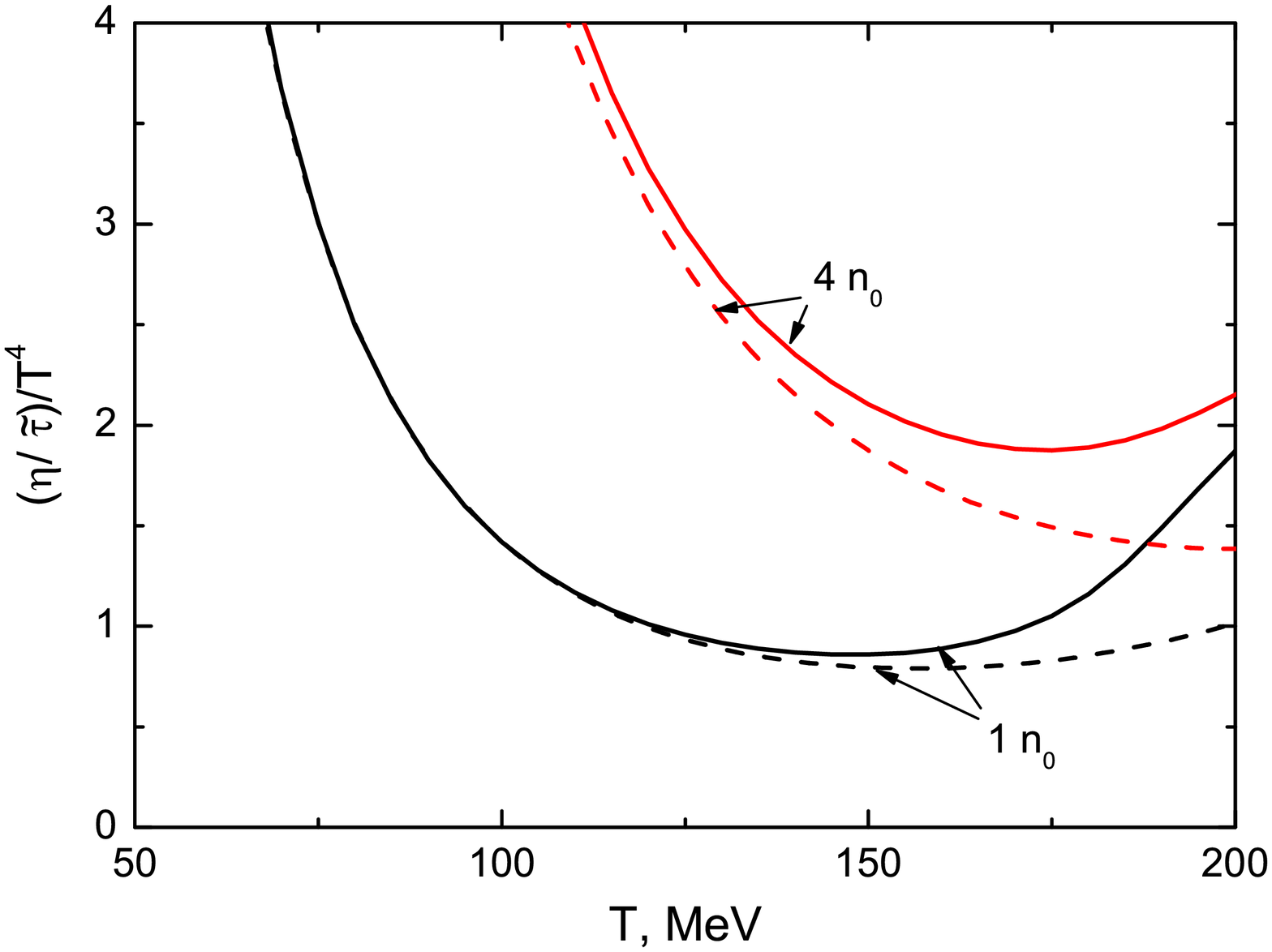}
\includegraphics[width=65mm,clip]{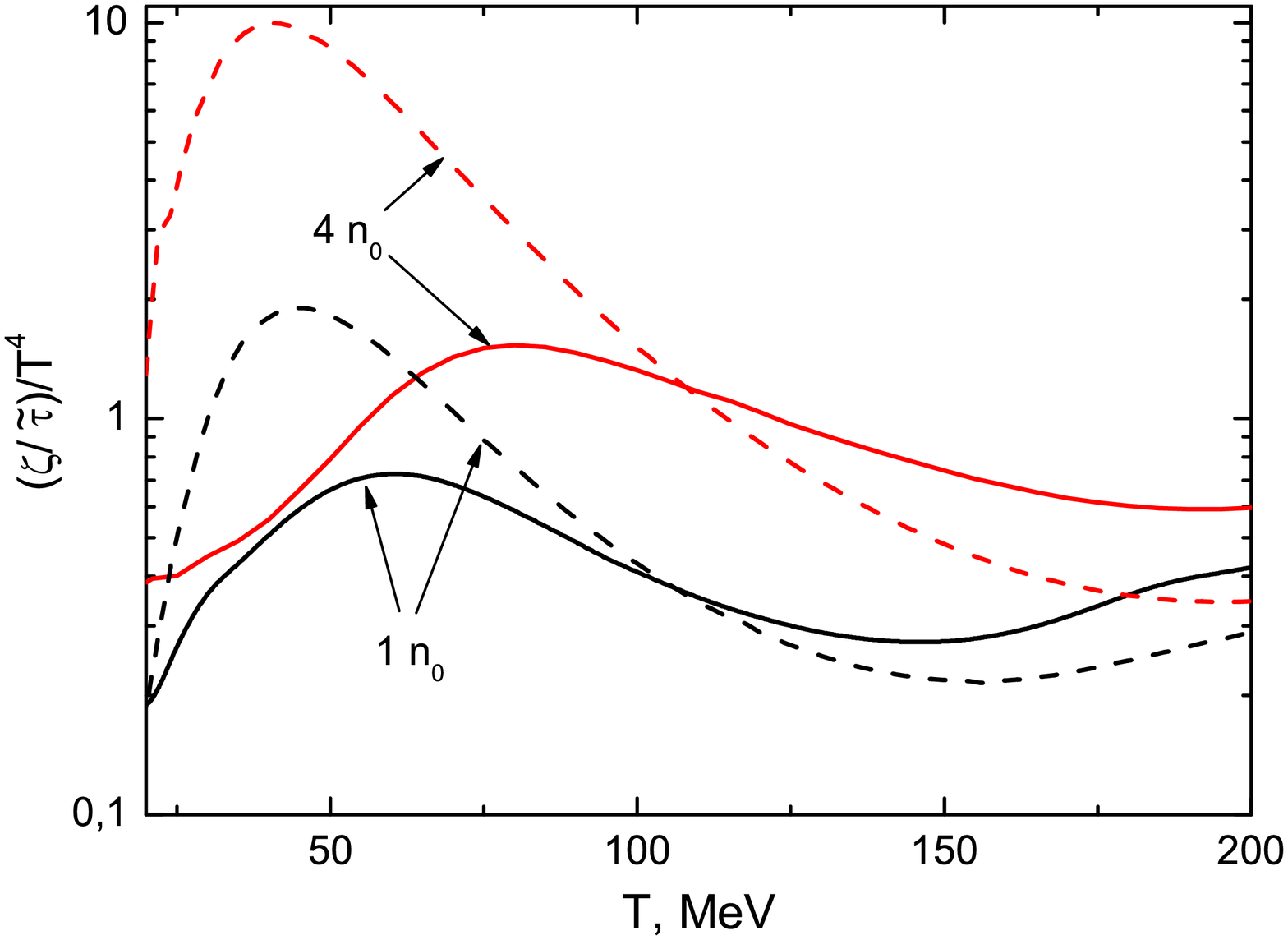}
\caption{ The SHMC  model predictions of the $T^4$-scaled
temperature dependence of the reduced shear (left panel) and bulk
(right panel) viscosities calculated
 for  hadron mixture at  $n_{\rm bar}=n_0$ and  $4n_0$ (solid lines).
 Calculations performed in the IG based model
with the same hadron set as in the SHMC model are demonstrated by
dashed lines.
 }
 \label{eta_nB}
\end{figure}

 The temperature dependencies of
the reduced $T^4$-scaled shear and bulk viscosities calculated
 for the case of the multi-component hadron
mixture within IG and SHMC models are shown in the left and right
panels of Fig.~\ref{eta_nB} at baryon densities $n_{\rm bar}=n_0$
and  $4n_0$. The reduced shear viscosity calculated in the SHMC
model (solid lines) is close to that in the IG model with the same
hadron set (dashed lines). Differences in the $\eta/(\tau T^4)$
ratio for the IG and SHMC models   appear only at high
temperatures $T\gsim 150$ MeV. At $T\lsim  100$ MeV the reduced
$T^4$-scaled bulk viscosity (right panel) in the IG based model
proved to be  larger than that in the SHMC model. Contrary, for
larger $T$ the values of the reduced bulk viscosity in the IG
model become smaller than those in the SHMC model. Differences
come from the strong dependence of the bulk viscosity $\zeta$ on
the values of thermodynamical quantities (see Eqs. (\ref{Qa}),
(\ref{bulk})).  Note that at $T\gsim$100 MeV and $n_{\rm bar}\gsim
n_0$ the shear and bulk viscosities are getting comparable in
magnitude. Growth of the relative importance of $\zeta$ with
increase of temperature seems to be quite natural because the bulk
viscosity takes into account
 momentum dissipation due to inelastic channels and those number
grows with the temperature increase.

Now let us proceed to an estimate of the relaxation time for a
 baryon enriched nuclear system. Important peculiarity of the
nucleon contribution to the relaxation time at low temperature is
associated with the particular role played by the Pauli blocking.
It means that appropriate multi-dimensional integration should be
carried out quite accurately with using quantum statistical
distribution functions. Calculations using the kinetic
Uehling-Uhlenbeck equations  for the purely nucleon system ($N=Z$)
in the non-relativistic approximation were performed
in~Ref.\cite{Dan84}.  For $T\lsim 100$ MeV an interpolated expression
has been obtained:
\be
{\tilde\tau}_{NN} \  &\simeq& \frac{850}{T^2} \left(\frac{n_{\rm
bar}}{n_0}\right)^{1/3} \ \left[1+0.04T \frac{n_{\rm
bar}}{n_0}\right] + \frac{38}{T^{1/2}(1+160/T^2)}\frac{n_0}{n_{\rm
bar}}~. \label{apptauN}
 \ee
Here ${\tilde\tau}_{NN}$ is in fm$/c$, $T$ is in MeV, $n_0 =0.145$
fm$^{-3}$. Thus the relaxation time demonstrates well known
$T^{-2}$ behavior for a Fermi liquid at $T\to 0$,
cf.~\cite{Tom38}.

\begin{figure}[t]
 \hspace*{20mm}
\includegraphics[width=100mm,clip]{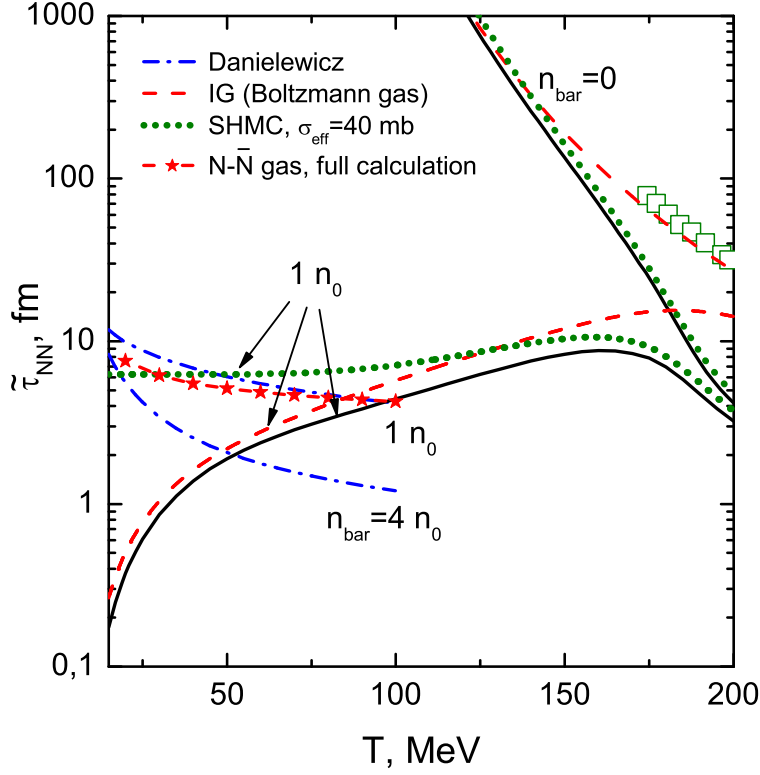}
\caption{ The temperature dependence of the partial
nucleon-nucleon relaxation time for $n_{\rm bar}=0$ and $1n_{0}$.
 SHMC model calculations with the relativistic Boltzmann
distribution function performed following (\ref{tau}) with the
experimental free nucleon-nucleon cross section are shown by solid
lines   and those with the effective cross-section
$\sigma_{NN}=\sigma_{eff}=$40 mb, by dotted lines, respectively.
The full calculation for nucleon component at $n=n_0$ performed
with the Fermi distribution function is presented by stars.
 The ideal Boltzmann nucleon-antinucleon gas  model results
are shown by dashed curves. The Danielewicz interpolation given by Eq.
(\ref{apptauN}) of~\cite{Dan84} is plotted by dash-dotted lines
for  $1n_{0}$ and   $4n_{0}$. Squares are results of calculations
~\cite{PPVW93} for $n_{\rm bar}=0$ performed within the Boltzmann
approximation. }
 \label{tau_nB}
\end{figure}

In Fig.\ref{tau_nB} we compare evaluation (\ref{apptauN}) with the
relaxation time, as it follows from the SHMC model for the purely
nucleon-antinucleon system with $N=Z$. Here we use the Boltzmann
approximation and free cross sections just corrected by the
nucleon mass shift (see solid lines). As is seen from
Fig.\ref{tau_nB}, due to an account for the Pauli principle  the
Danielewicz's ${\tilde\tau}_{NN}$ given by Eq. (\ref{apptauN})
 (see dash-dotted line)  essentially deviates from the SHMC model
result (computed here within Boltzmann approximation) for $T\lsim$
100 MeV. The SHMC model results are very close to those for the
ideal nucleon-antinucleon gas (dashed lines) for $T\lsim 150$ MeV.
A partial account of the Pauli principle by assuming constant
effective cross section $\sigma_{eff} =$40 mb (see dotted line)
improves SHMC model agreement with the estimates (\ref{apptauN})
at low $T$. Our full calculation of ${\tilde\tau}_{NN}$ for a
purely nucleon gas that includes Pauli principle plotted by stars
demonstrates a reasonable agreement with (\ref{apptauN}).
Therefore to simplify  calculations below we use Eq.
(\ref{apptauN}) for the partial nucleon-nucleon relaxation time
$\tilde\tau_{NN}$, to be valid at low temperatures,  smoothly
matching it (at $T\sim 100$ MeV)  with the partial nucleon
contribution calculated following Eq. (\ref{tau}) for higher
temperatures. In the case $n_{\rm bar}=$0 the ideal Boltzmann  gas
model results (dash curve) are very close to those of
~\cite{PPVW93} (squares) and agree with the SHMC calculations
(also performed in the  Boltzmann approximation, see solid and
dotted lines) for $T\lsim 150$MeV. For higher $T$ the difference
with the IG model is due to a sharp change of the effective
nucleon mass in the SHMC model at large temperatures.

In subsequent calculations we take into account the whole hadron
set involved into the SHMC model. The relaxation time for every
component, except for nucleons, is evaluated according to Eq.
(\ref{tau}).
\begin{figure}[thb]
 \bc
\includegraphics[width=120mm,clip]{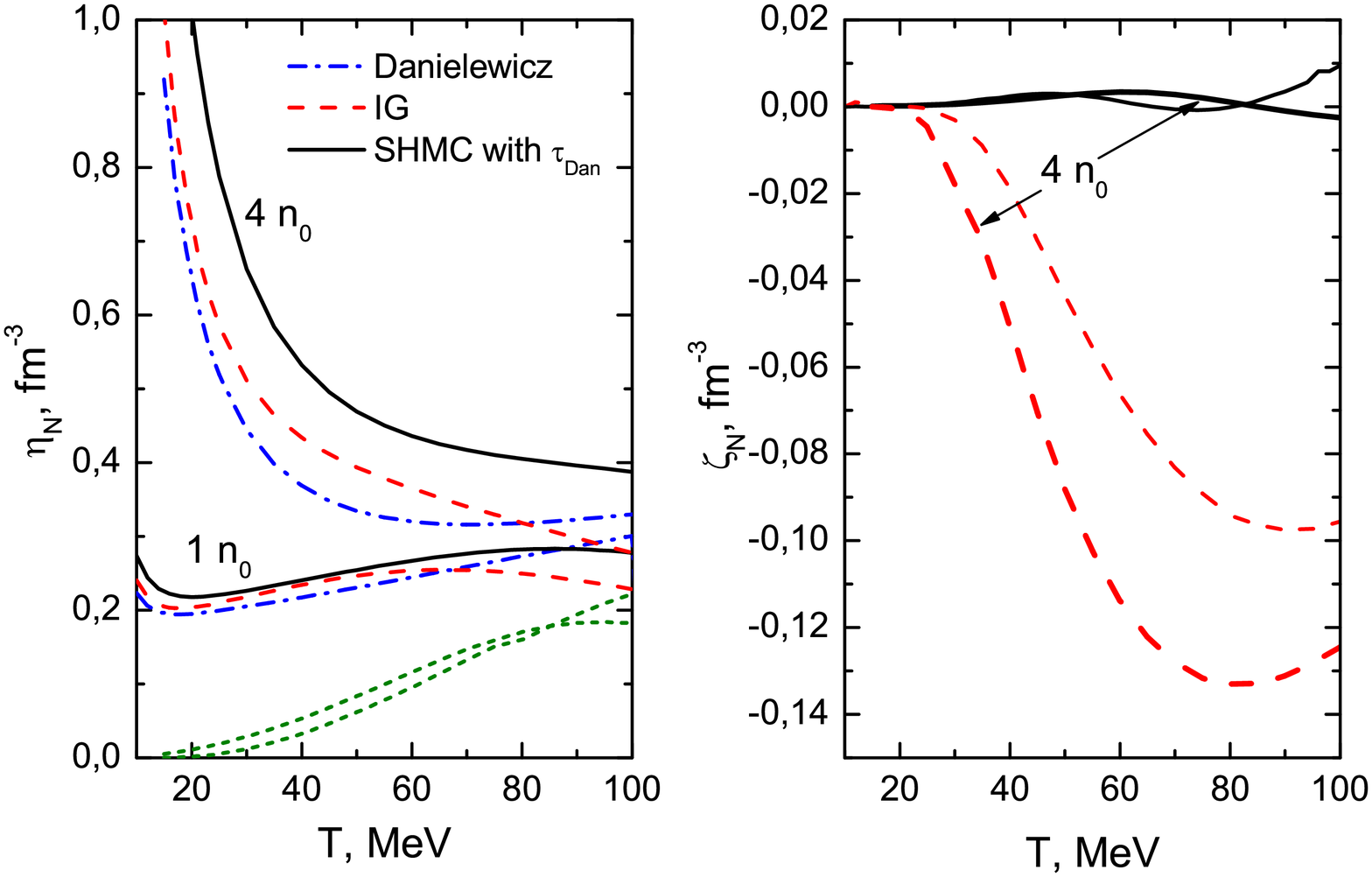}
\caption{ The $T$-dependence of the partial nucleon contribution
to the shear and bulk viscosity in the multi-component system  for
different values of the baryon density $n_{\rm bar}$. Solid and
dashed lines are  our results, respectively, for the SHMC and IG
models with the same particle set,  with the relaxation time
calculated following Eq. (\ref{apptauN}).
 Dot-dashed curves are the Danielewicz interpolation
(\ref{appD}) of  results~\cite{Dan84}. For comparison in left
panel two short-dashed lines are shown for the SHMC model
calculations done within the  Boltzmann gas approximation  at
$n_{\rm bar}=4n_0$ (upper line in low $T$ region) and $n_0$. In
right panel bold solid and dashed lines correspond to $n_{\rm
bar}=4n_0$ and thin ones, to $n_{\rm bar}=n_0$.}
 \label{etaD_nB}\ec
\end{figure}

 In Fig.\ref{etaD_nB} (left) the partial contribution  of the nucleon
shear viscosity for  the multi-component system is pictured as a
function of temperature at $n_{\rm bar}=1n_0$ and 4$n_0$. Solid
and dashed lines are our results for the SHMC and IG models,
respectively, with the relaxation time calculated following Eq.
(\ref{apptauN}). The dot-dashed lines show  the Danielewicz analytical
fit~\cite{Dan84}
\be
\eta \simeq &&(1700/T^2) \ (n_{\rm bar}/n_0)^2+[22/(1+T^2\cdot
10^{-3})] \ (n_{\rm bar}/n_0)^{0.70} \nonumber
\\&+&5.8T^{1/2}/(1+160/T^2)~,
 \label{appD}
  \ee
  where $\eta$ is
given in MeV/(fm$^2 c)$, $T$ in MeV. For  both $n_{\rm bar}/n_0=$1
and 4 the IG model results are rather close to this fit, provided
$\tilde{\tau}_{NN}$ is calculated according to (\ref{apptauN}).
The shear viscosity calculated under the same conditions within
the SHMC model is higher than that calculated following Eq.
(\ref{appD}) and in IG model, since the former takes into account
decrease of the effective masses with the increase of the baryon
density.
 Two short-dashed curves at the bottom of
Fig.\ref{etaD_nB} (left) represent $\eta$ computed in the SHMC
model within the Boltzmann  approximation for densities $n_{\rm
bar}=n_0$ and $4n_0$. As is expected, in the Boltzmann case at low
temperatures the shear viscosity only weakly depends on the baryon
density
(cf. short-dashed curves in Fig.\ref{etaD_nB}).
This has been noted earlier in~\cite{Gal79}. Our results
demonstrate important role played by the Pauli principle at low
$T$ and show a strong density dependence.
\begin{figure}[thb]
\bc
\includegraphics[width=120mm,clip]{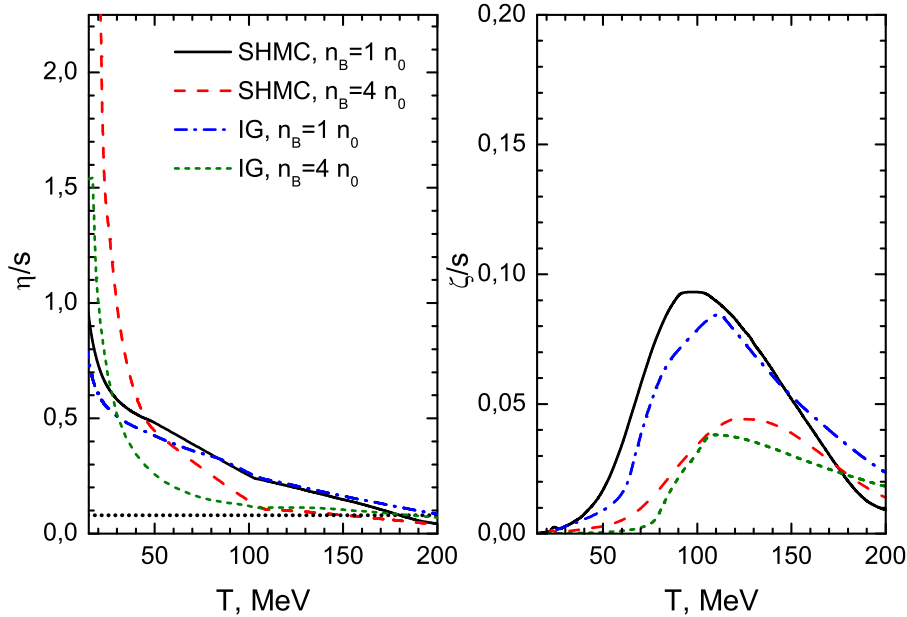}
\caption{ The $T$-dependence of the specific viscosity
coefficients $\eta/s$ and $\zeta/s$ for two values of the baryon
density $n_{\rm bar}$. Solid and dashed lines are our results for
the SHMC model for $n_0$ and $4n_0$, respectively. The dash-dotted
and dashed lines are calculations within the IG model with the
same hadron set, again for $n_0$ and $4n_0$. Relaxation time is
calculated following Eq. (\ref{apptauN}). The dotted line
corresponds the the lower AdS/CFT bound $\eta/s=\frac{1}{4\pi}$.}
 \label{etaz_nB}\ec
\end{figure}

The model dependence of results is more pronounced for $\zeta$.
 In
the right panel of Fig.\ref{etaD_nB}  we present the partial
contribution of the nucleon bulk viscosity for the multicomponent
system. We see that in the SHMC model this contribution is very
small and can be neglected. In the IG model the partial
contribution, $\zeta_N$, proves to be negative, whereas the total
value $\zeta$ remains, certainly, positive.

The temperature dependence of the specific $\eta/s$ and $\zeta/s$
transport coefficients is presented in Fig.\ref{etaz_nB} for two
values of the baryon density $1n_0$ and $4n_0$. The results of the
SHMC (solid and dashed lines)  and IG (dash-dotted and
short-dashed lines) models differ not very much for the given
value of the baryon density, except for  the shear viscosity at
low temperature due to the Pauli exclusion principle. The $\eta/s$
for the baryon rich matter achieves  the lower AdS/CFT bound
$\eta/s=\frac{1}{4\pi}$
 at smaller $T<T_c$ for the higher $n_{\rm bar}$.

 \subsection{Collisional viscosity in heavy ion collisions}

Above we have studied the specific viscosities of the hadron matter
at different temperatures and baryon densities. In reality a hot
and dense system formed in a heavy-ion collision
expands toward a freeze-out state, at which the components stop to
interact with each other.

 We should specially note that the approximations of the slow
hydrodynamic expansion are violated at the freeze-out stage, which
 is  assumed to be instantaneous
 within simple hydrodynamical models. More elaborated approach
assuming continuous freeze-out decoupling demands some hybrid of
kinetic and hydrodynamic description e.g.,
see~\cite{Arsene,Kn09}. Rapid processes out of equilibrium may lead to an additional
dissipation and particle and entropy productions
\cite{VS89,Magas}. These effects are beyond the scope of the
relaxation time approximation to the quasiparticle Boltzmann
equation used in this work to evaluate the collisional viscosity.
 In addition,  there are other sources contributing to
viscosity, see e.g. Ref.~\cite{PP06} and also Appendix C.
 Here we use the phenomenological
freeze-out curve $T_{\rm fr}(\mu_{\rm bar}^{\rm fr})$ extracted
from the analysis of experimental particle ratios in statistical
model for many species at the given collision energy
$s_{NN}^{1/2}$ treating  the freeze-out temperature $T_{\rm fr}$
and chemical potential $\mu_{\rm bar}^{\rm fr}$ as free
parameters~\cite{COR06,ABMS06}. Thereby extra particle production is phenomenologically
incorporated. We may also hope that including particle production
effects we also partially include the corresponding effects of the
entropy and viscosity productions.

\begin{figure}[thb]
 \hspace*{20mm}
\includegraphics[width=90mm,clip]{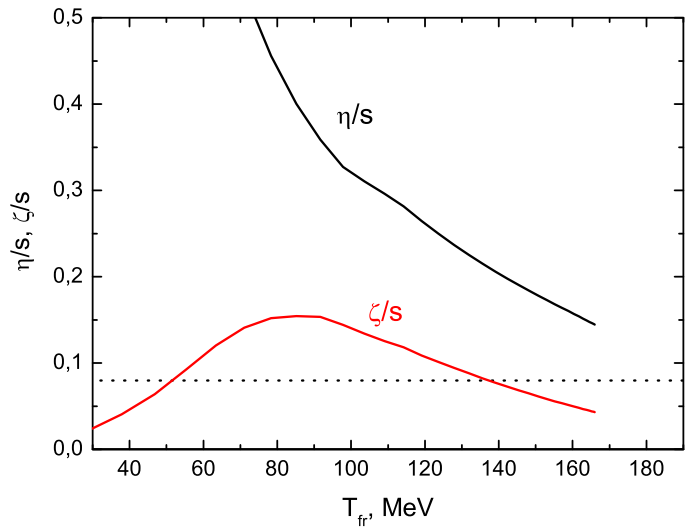}
\caption{Specific shear and bulk viscosities  calculated
in the SHMC model  for central Au+Au collisions along the
freeze-out curve (at $T=T_{\rm fr}$) ~\cite{ABMS06} for the baryon
enriched system. The dotted line is the lower AdS/CFT bound
$\eta/s=1/4\pi$~\cite{KSS03}.
 }
 \label{sh_b_fr}
\end{figure}

 In Fig.~\ref{sh_b_fr}, $\eta/s$, $\zeta/s$ ratios  are shown for Au $+$
Au collisions  versus
 the freeze-out temperature (which  is unambiguously related to the freeze-out
chemical  potential $\mu_{\rm bar}^{\rm fr}$~\cite{ABMS06} needed
to calculate thermodynamical quantities at the freeze-out).
 As  we see,  the $\eta/s$ ratio decreases monotonously
with increase of the temperature, being higher than the lower
bound $1/4\pi$ but tending to it with further increase of $T_{\rm
fr}$.  The value $\zeta/s$ exhibits a maximum at $T_{\rm fr}\sim
85$ MeV and then tends to zero with subsequent increase of $T_{\rm
fr}$. As has been emphasized above, at $T\gsim$100 MeV values of
the shear and bulk viscosities become  comparable in value,
$(\eta/s)_{\rm fr}\simeq 2(\zeta/s)_{\rm fr}$.
\begin{figure}[thb]
 \hspace*{20mm}
\includegraphics[width=90mm,clip]{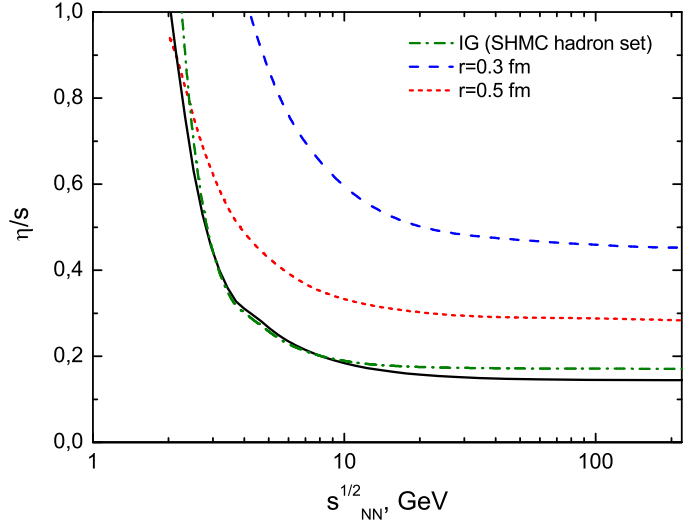}
\caption{The specific shear  viscosity
calculated for central Au+Au collisions along the chemical
freeze-out curve~\cite{ABMS06} within the SHMC model as a function
of the collision energy $s^{1/2}_{NN}$ (solid line). Dashed and
short-dashed curves are the results of the excluded volume hadron
gas model~\cite{GHM08} with hard-core radii $r=$0.3 and $r=$0.5
fm, respectively. The dot-dashed line corresponds to the  IG model
with the same set of hadrons as for the SHMC model. }
 \label{sh_Gfr}
\end{figure}

 In Fig. \ref{sh_Gfr}, the $\eta/s$ ratio calculated in our SHMC
model  (solid line) is plotted as a function of the collision
energy $\sqrt{s_{NN}}$ of two  Au+Au nuclei. The result for the IG
model with the same hadron set as in SHMC model  is plotted by the
dash-dotted line. We  note that for $\sqrt{s_{NN}}\gsim 3$ the
SHMC results prove to be very close to the IG based model ones
(with the same hadron set as in SHMC model), since the freeze-out
density is rather small and the decrease of the hadron masses
occurring in the SHMC model is not important. The results for  the
hadron hard core gas model (the van der Waals excluded volume
model)~\cite{GHM08} at two values of the particle hard core radius
$r$  are shown by dashed and short-dashed lines. In all cases for
$\sqrt{s_{NN}}\gsim 2$ GeV the ratio $\eta/s$ decreases along the
chemical freeze-out line with increasing  collision energy and
then flattens at $\sqrt{s_{NN}}\gsim $10 GeV, since  freeze-out at
such high collision energies already occurs at almost constant
value of $T_{\rm fr}\approx 165$ MeV. The shear viscosity  of the
non-relativistic Boltzmann gas of hard-core particles~\cite{GHM08}
is $\propto{\sqrt{mT}}/{r^2}$. Since Fermi statistical effects are
not included within this model, the shear viscosity, $\eta$,
decreases with decrease of $T$. Nevertheless the $\eta/s$ ratio
increases and diverges at low energy/temperature, as the
consequence of a more sharp decrease of the entropy density
compared to $\eta$, see Fig.~\ref{sh_Gfr}. As follows from the
figure,  the smaller $r$ is, the higher $\eta/s$  is in the given
excluded volume model. For $\sqrt{s_{NN}}\gsim 4$ and $r\simeq
0.7$ fm the $\eta/s$ ratio is expected  to be close
to the values computed in the IG and SHMC
 models. We also note that in the whole range of the considered
$\sqrt{s_{NN}}$ the SHMC results proved to be very close to the
IG-based model expectation (with the same hadron set as in SHMC
model) since the freeze-out density is small.

Recently an interesting attempt has been undertaken in~\cite{IMRS09}
to extract the shear viscosity from the 3-fluid hydrodynamical analysis
of the elliptic flow in the AGS-SPS energy range. An overestimation
of experimental elliptic flow $v_2$ values obtained in this
model was associated with dissipative effects occurring during the
expansion and freeze-out stages of participant matter evolution.
 The resulting
values of $\eta/s$  vary in  interval $\eta/s\sim 1-2$ in the
considered domain of $\sqrt{s_{NN}}\approx 4-17$ GeV
(corresponding to temperatures $T\approx 100-115$ MeV). Authors
consider their result as an upper bound on the $\eta/s$ ratio in
the given energy range.
 Note that the mentioned values are much higher than
those which follow from our estimations given above and presented
in Figs. \ref{sh_b_fr} and \ref{sh_Gfr}.  One should
additionally mention that the local equilibrium in the three-fluid models
 is established in every fluid but not among fluids in a common cell,
whereas other
models considered above imply a local thermal equilibrium.

Other microscopic estimate of the share viscosity to the entropy
density ratio for the relativistic hadron  gas based on the Kubo formulae
was performed in Ref.~\cite{DB08}. Calculations were carried out
 by simulation of a system evolution in a box with appropriare
 boundary conditions using the UrQMD code,
where 55 baryon species and
their antiparticles and 32 meson species were included. The full
kinetic and chemical equilibrium is achieved at $T=$130 and 160
MeV, respectively. Treating these temperatures as kinetic and
chemical freeze-out ones and comparing them with those in Fig.
\ref{sh_b_fr}, we see that the extracted ratio $\eta/s\gsim$1
exceeds the SHMC result by a factor of  5. Introducing a non-unit
fugacity or a finite baryon density allows one to decrease the
ratio twice but nevertheless it is still too high as compared to
both the SHMC result and  the lower bound $\eta/s=1/4\pi$.
Analyzing their result  authors~\cite{DB08} conclude that the
dynamics of the evolution of a collision at RHIC is  dominated by
the deconfined phase (exhibiting very low values of $\eta/s$)
rather than by the hadron phase. Note however that in-medium
effects in the hadron phase are not included into consideration in
the UrQMD model though, namely, these effects result in the
required decrease of the $\eta/s$ ratio in our SHMC model.

Similar dynamical estimate has been done recently in
Ref.~\cite{Pal10} where additionally to real hadrons the Hagedorn
states were considered. Inclusion of heavy Hagedorn clusters
decreases the $\eta/s$ ratio at $T\lsim$140 MeV as compared to the
UrQMD result~\cite{DB08} cited above but the obtained value
$\eta/s\approx 0.3-0.4$ is still higher than the lower AdS/CFT
bound and the Hagedorn gas statistical estimate for $\mu_{\rm
bar}=$0~\cite{NHNG08} (see Fig. \ref{sh_b_T}). It is worthy to
note that for a finite  chemical potential (or baryon density) the
specific shear viscosity in~\cite{Pal10}  is getting lower  but in
contrast with our SHMC model (see Fig. \ref{etaz_nB}) the lower
AdS/CFT bound is not reached.

\section{Shear and bulk viscosities in the quark
phase}\label{Quark}

\subsection{Estimates of the relaxation time}

In the framework of the quasiparticle approximation the shear and
bulk viscosities in the QGP phase can be found with the help of
Eqs.~(\ref{etaC}), (\ref{zetaC}). These equations should be
 supplemented by expressions for the collisional relaxation time.  Like
in  the hadron system, the  collisional relaxation time $\tau_a$,
$a=\{q,g\}$, depends on the momentum. To simplify our calculations
of the transport coefficients we will use for  $\tau$  values
${\tilde\tau}_a$, where  ${\tilde\tau}_a$ is estimated with the
thermal averaged cross-sections describing total elastic
scattering of medium constituents. Unlike the hadronic cross
section the quark/gluon (parton) elastic scattering cross section
is not measurable and should be evaluated in a model. Really, it
is a specific problem. The in-medium cross sections for
quark-antiquark, quark-quark and antiquark-antiquark scattering
processes were studied in detail in Ref.~\cite{ZKHKn95} within the
NJL model  for two different flavors, including $1/N_c$
next-to-leading order corrections. These results incorporate
dominance of the scattering on  large angles and take into account
a possible occupation of particles in the final state, see
Ref.~\cite{SR08c}.

The QCD calculations of  the relaxation time ${\tilde\tau}$ of
partons already in the lowest order in  the running coupling
constant $g$ require summation of infinitely many diagrams.
Resummation of the hard thermal loops results in the width
${\tilde\tau}^{-1}$ of partons $\sim g^2T \ln (1/g)$ \cite{Pi89}.
Thus, the following parametrization was used for
gluons~\cite{PC05,Pe01}
 \be
{\tilde\tau}_g^{-1}=2N_c \frac{g^2T}{8\pi} \ln \frac{2c}{g^2},
 \label{tau_g}
 \ee
and similarly for quarks
 \be
{\tilde\tau}_q^{-1}=2\frac{N_c^2-1}{2N_c} \frac{g^2T}{8\pi} \ln
\frac{2c}{g^2},
 \label{tau_q}
 \ee
where the running coupling constant is given by
 \be g^2(T)=\frac{48\pi^2}{(11N_c-2N_f)\ln
(\lambda(T-T_s)/T_c)^2}~,
 \label{g2T}
  \ee
which permits an enhancement near $T_c$ (by construction,  the
hadron-quark phase transition in this model is of the first
order). Parameters of the effective coupling $T_s$ and $\lambda/
T_c$ were adjusted to the lattice QCD EoS similar to what we
did above for the two-phase EoS.  One uses $T_s/T_c=$0.46 and
the tuning parameter $c$ is determined from the condition
$g^2(T=T_c)=0$ ~\cite{PC05,Cassing} that results in the divergence
of the relaxation time at $T_c$ (the cross section is zero at this
point).
\begin{figure}[thb]
 \hspace*{20mm}
\includegraphics[width=90mm,clip]{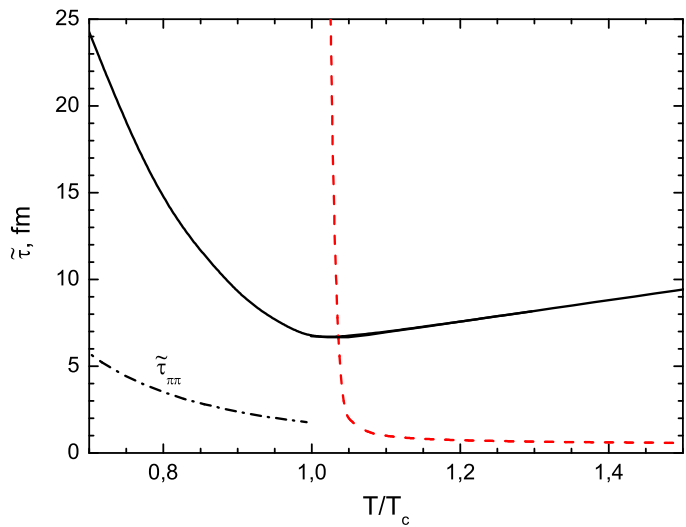}
\caption{  Temperature dependence of the relaxation time
$\widetilde{\tau}_q$ for $\mu_{\rm bar}=0$.   The solid line shows
result of a calculation  within the NJL model~\cite{SR08c}. The
quark relaxation time $\widetilde{\tau}_q$
 estimated according to Eq. (\ref{tau_q}) at $T/T_c
>1$ is given by the dash line. For comparison the relaxation time
$\widetilde{\tau}_{\pi\pi}$ obtained following Eq. (\ref{tau}),
but only for the gas of massive free pions at $T/T_c<1$, is
plotted by the dash-dotted line.
 }
 \label{tau_gl}
\end{figure}

The temperature dependence of ${\tilde\tau}_q$ for $\mu_{\rm
bar}=0$ within  the NJL model~\cite{SR08c} is shown in
Fig.~\ref{tau_gl}  by the solid line. Both $T<T_c$ and $T>T_c$
regions are covered in this model.  However  the NJL model, in
which the medium is assumed to be composed only of heavy
constituent quarks, can hardly be applied for the description of
the $T<T_c$ phase. Therefore at $T<T_c$ for this phase we suggest
to use the developed above hadronic description, where collisional
time is calculated following Eq. (\ref{tau}) with experimental
cross-sections for hadronic species. In the case $\mu_{\rm bar}=0$
the main contribution is given by the pion term (see dash-dotted
curve). In  Fig. \ref{tau_gl} we also  show the relaxation time
$\widetilde{\tau_q}$  for $T>T_c$ following (\ref{tau_q}), see the
dashed line in figure.   In order to obtain the relaxation time
for gluons, $\widetilde{\tau}_g$, we should multiply
$\widetilde{\tau}_q$ by the group factor $4/9$.

\subsection{Collisional viscosity in the quark-gluon phase}

 In the large $N_c$ limit the specific shear viscosity in the QGP
phase can be inferred from calculations of \cite{K08}
\be
\left(\frac{\eta}{s}\right)_{QGP}=\left(\frac{1+3.974\xi}{1+1.75\xi}\right)
\frac{69.2}{(g^2N_c)^2\ln (26/(g^2N_c(1+0.5\xi))} \label{esQGP}
\ee
 with $\xi=N_f/N_c$.
Obviously this ratio  in the quark/gluon phase at
 finite large $N_c$  differs
from that calculated for the hadronic phase. Therefore, one may
expect a jump in the temperature dependence of the $\eta/s$-ratio
at the crossing of the phase boundary of the first-order phase
transition.
\begin{figure}[thb]
 \hspace*{2mm}
\includegraphics[width=140mm,clip]{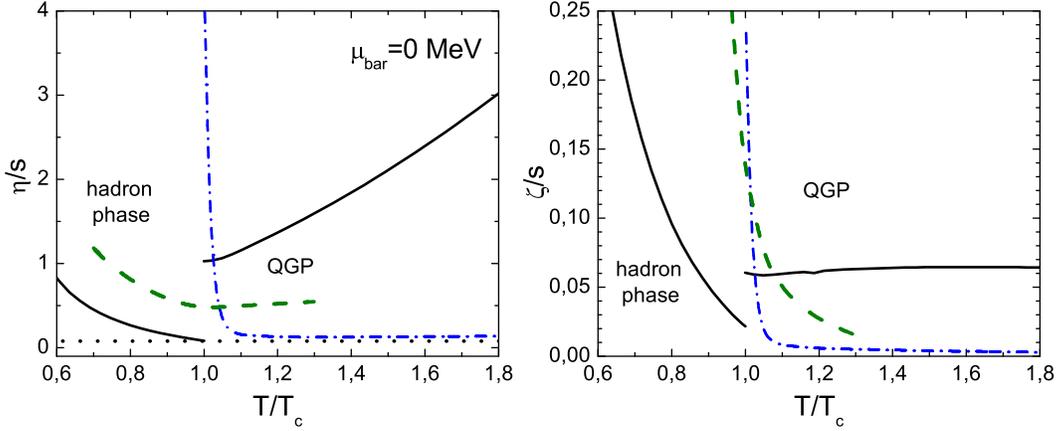}
\caption{ The $T$-dependence of the shear (left panel) and bulk
(right panel) specific viscosities within our two-phase SHMC-HQB
model (solid lines) for $\mu_{\rm bar}=$0. The results for the
original NJL model~\cite{SR08c} are plotted by dashed lines. The
$\eta/s$ and $\zeta/s$ calculated following Eqs.~(\ref{etaC}),
(\ref{zetaC}) using quark masses from the HQB model and the
relaxation time from (\ref{tau_q}) are shown by the dash-dotted
lines. The lower bound for the reduced shear viscosity, $1/4\pi$,
 is given by the horizontal dotted line.
 }
 \label{sh_b_2PH}
\end{figure}
 In our HQB
model strange quarks and gluons are very massive. Therefore their
contribution to transport coefficients and  entropy density
is minor. Main contributions arise from $u$- and $d$-quarks.

In Fig.~\ref{sh_b_2PH}, $\eta/s$ and $\zeta/s$ for hadron
 and quark-gluon phases are shown in a broad temperature range
 for $\mu_{\rm bar}=0$. Solid lines present results of the two-phase
SHMC-HQB model. Below $T_c$ the ratios are  the same as shown
above in Fig.~\ref{sh_b_T}.  Calculations done in the NJL
model~\cite{SR08c} are shown by dashed lines.  In both cases
the quark collision relaxation time
is that shown in Fig.~\ref{tau_gl} by the solid line.  We see that the
original NJL model~\cite{SR08c} gives a continuous smooth line (a
crossover) for $\mu_{\rm bar}=0$. In our SHMC-HQB model there is a
jump at $T_c$ in both the $\eta/s$ and $\zeta/s$ ratios.
This jump is a particular property of the first order phase
transition. For $\mu_{\rm bar}=$0 in our two-phase SHMC-HQB model
the $\eta/s$ ratio  increases with $T$ more sharply  after the
jump has occurred compared to that in the original NJL model (see
Fig.~\ref{sh_b_2PH}).

Dash-dotted curves in Fig.~\ref{sh_b_2PH} demonstrate that the
specific viscosities diverge for $T\rightarrow T_c +0$ since
$\widetilde{\tau} \to \infty$ provided the nonperturbative
estimate of the relaxation time (\ref{tau_q}) is used. With
increase  of the temperature above $T_c$ the temperature
dependence of these ratios  flattens.   The gluon $\eta/s$ and
$\zeta/s$ ratios exhibit in this case a similar divergent behavior
near $T_c$, since the relaxation time (\ref{tau_g}) also diverges
at $T_c +0$. As follows from Eq. (\ref{esQGP}), in the QGP phase
$\eta/s \propto 1/g^4$ where in accordance with (\ref{g2T}) the
coupling constant is expressed via $g^2\propto 1/\ln
(T/\Lambda_T)$,  $\Lambda_T$ is proportional to the scale
parameter of QCD. Similar minimum close to $T_c$ from the QGP side
 arises for the $\eta/s$
for the gluon quasi-particle excitations in a phenomenological
quasiparticle model~\cite{BKR09}  in agreement with lattice QCD
data in gluodynamics~\cite{NS05,SN07,Mey07}. Thus we conclude that
both the magnitude of the jump and the $\eta/s$  behavior for
$T>T_c$
are essentially model-dependent. Generally, similar conclusion can
be done from comparison of different curves for $\zeta/s$ in the
right panel of Fig.~\ref{sh_b_2PH}.

Note  that for the hadron phase the NJL model noticeably
overshoots the SHMC model results for both $\eta/s$ and $\zeta/s$
specific viscosities. However one should be careful comparing the
NJL  and SHMC-HQB results. The NJL model treats the chiral phase
transition while in the SHMC-HQB model we deal with deconfinement.
Thus, critical temperatures have different meaning in these
models. Their values are also different (but  the latter problem
can be avoided provided one uses $T/T_c$ as an argument).
 Moreover in the NJL model one continues to deal  only with $u$ and
$d$ quark degrees of freedom at $T<T_c$, whereas  the SHMC model
involves many hadron degrees of freedom whose masses essentially
decrease at $T\sim T_c$. Thereby, the entropy density in the SHMC
model is essentially higher and the specific viscosity is
accordingly lower than those in the NJL model~\cite{SR08c}. The
shear viscosity-to-entropy ratio in the latter model is much
higher than its lower bound $1/4\pi$ at $T\approx T_c$.

One should note that in our model the deconfined partonic phase is
described in a simplified way as an ideal massive gas but
relaxation times were taken from the models including
interaction. A consistent approach requires an account for
interactions between the constituents  and the extracted effective
couplings should enter the estimate for the transport cross
section. A step toward this direction was undertaken in
Ref.~\cite{MC09} within the generalized classical virial expansion
formalism. The corrections to a single particle partition function
were calculated starting from a parton interaction potential whose
parameters are fixed by lattice thermodynamical quantities.
The dependence of the $\eta/s$ ratio on the temperature dependence
of the strong interaction coupling has been studied. Results prove
to be sensitive to the choice of the interaction.

There exist many  other model estimates of the $\eta/s$ ratio for
$T>T_c$  for $\mu_{\rm bar}=$0.  As noted in the Introduction,
AdS/CFT \cite{PSS01} predicts $\eta/s=1/(4\pi)$. Lattice Monte
Carlo calculations are available only for gluodynamics. For the
SU(3) gauge theory a robust upper bound $\eta /s< 1$
 was obtained with the best estimate
$\eta/s=0.134$ at $T=1.65 T_c$~\cite{Mey07}. A slightly larger
value of the shear viscosity was derived in the lattice QCD
simulation in the quench approximation for  a hot gluon
gas~\cite{NS05}: $\eta/s \simeq 0.1\div 0.4$ for temperature in
the region $1.4 \leq T_c\leq 1.8$. At higher temperatures $\eta$
increases by two-three orders of magnitude.

\begin{figure}[thb]
 \hspace*{2mm}
\includegraphics[width=140mm,clip]{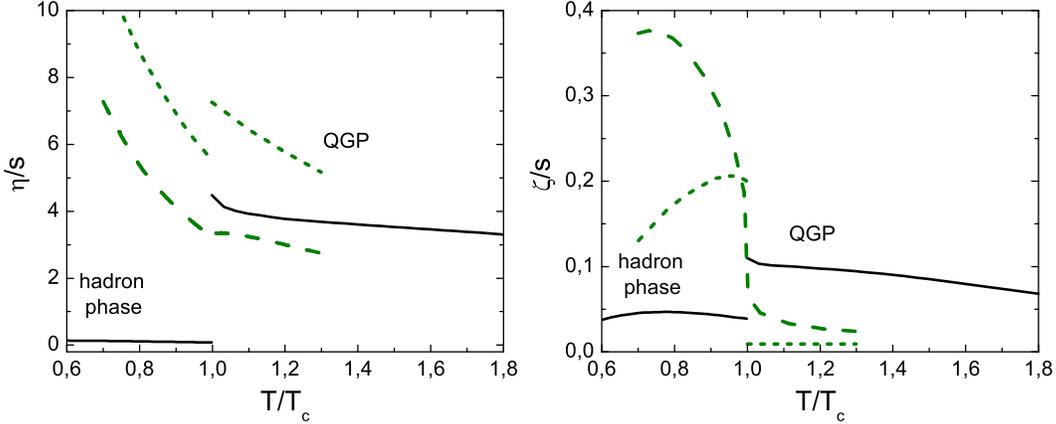}
\caption{ The $T$-dependence of the shear (left panel) and bulk
(right panel) specific viscosities within our two-phase SHMC-HQB model
(solid lines) for $\mu_{\rm bar}^{\rm CEP}=$990 MeV, corresponding
to the critical end point. The NJL model results~\cite{SR08c} for
$\mu_{\rm bar}^{\rm CEP}$ and for $\mu_{\rm bar}$ slightly above
$\mu_{\rm bar}^{\rm CEP}$
 are plotted by long-dash and short-dash
lines, respectively.
 }
 \label{sh_mu_2PH}
\end{figure}

 Model predictions for the $\eta/s$ and $\zeta/s$ ratios
 for systems with $\mu_{\rm bar}\ne 0$ are presented in
Fig.\ref{sh_mu_2PH}. The NJL model~\cite{SR08c} at the critical
end point ($\mu_{\rm bar}=$990 MeV) shows a smooth curve
(long-dash line)  with a sharp bend at $T_c$ for $\eta/s$ and
$\zeta/s$. Certainly, our two-phase model again exhibits a jump at
$T_c$
(see solid
lines). Similar jump will appear in the NJL model if $\mu_{\rm
bar}$ is larger than 990 MeV, {\it i.e.} when the system reaches
the region of the first order phase transition (see short-dashed
lines).

 Here the following remark
is in order: As  follows from all above estimates, the bulk
viscosity is smaller (but not much smaller in the hadron phase at
finite baryon density) than the shear one in the whole temperature
interval, may be except the vicinity of the critical point. In the
latter region soft modes may  additionally contribute significantly
increasing the bulk viscosity.

In spite of a variety of existing models, none of them is reliable
 for the description of the vicinity of the critical point.
In this respect, it is highly desirable  to get experimental
information about viscosity realized in heavy ion collisions to
compare it with the model predictions.
 The shear viscosity is intimately related to the observed
elliptic flow $v_2$ and viscous hydrodynamics provides a tool to
extract $\eta/s$ from experimental data. However, calculation
uncertainties are significantly affected by details of the
initialization of the initial hydrodynamic state, behavior of the
bulk viscosity and the speed of sound near the quark-hadron phase
transition point, chemical composition and strong non-equilibrium
effects during the late kinetic stage of the evolution of the
hadronic state. From the study of transverse momentum fluctuations
at the RHIC energies one gets $\eta/s \simeq 0.08\div 0.3$  in
Ref.~\cite{GAA06}. Close estimates $\eta/s \simeq 0.08\div 0.3$
and $\eta/s \simeq 0.09\div 0.15$ were obtained from $v_2$
analysis in  Refs. \cite{Drescher} and \cite{Lacey}, respectively.
In its turn,  studies of the heavy quark energy loss and $\phi$
meson production yield $\eta/s \simeq 0.1\div 0.16$~\cite{Adare}
and $\eta/s \simeq 0.07\pm 0.003\pm 0.014$~\cite{Chaudhuri},
respectively. Recent comprehensive analysis of the  hydrodynamic
simulations~\cite{SH08} has established an upper limit
\be
\frac{\eta}{s}\left|_{\rm QGP} <5 \times \frac{1}{4\pi}\right.~,
\label{up_lim}
 \ee
 to be rather close to the low AdS/CFT bound.
Note that the presence  of $\zeta \neq 0$ can also result in a
suppression of the elliptic flow $v_2$ competing at $T$ near $T_c$
 with shear viscous effects \cite{SH09}. Thus, the above
constraint on $\eta/s$ can be even stronger.

 Finally, one should emphasize that practically in all
viscous hydrodynamic calculations the specific ratio  $\eta/s$  is
considered as a time-independent quantity,  that is  not the case,
as follows from the above-presented dependence of kinetic
coefficients on temperature and baryon density. In addition, it is
tacitly neglected by retardation effects in the formation of
irreversible current. Some exceptions
are  estimates~\cite{Boz09,DKK10}. In Ref.~\cite{Boz09} viscosity
effects were studied with two different $T$-dependent shear and
bulk viscosities (treated as free parameters) in the QGP and
hadronic phases. The values of the used transport coefficients
noticeably differ from those presented in  Figs.\ref{sh_b_2PH} and
\ref{sh_mu_2PH}. Following~\cite{Boz09}  the elliptic flow
coefficient is significantly reduced due to viscosity effects both
in the plasma and in the hadron matter. A more realistic parameterization
of the temperature dependence of shear and bulk
viscosity coefficients was considered in Ref.~\cite{DKK10} and
additionally relaxation times were included in order to explain
elliptic flow at RHIC. It was argued that close to ideal behavior
observed at RHIC energies may be related to a non-trivial temperature
dependence of these transport coefficients. So,
extraction of the shear viscosity from analysis of the initial
QGP phase is a difficult problem which can be reliably addressed
only after precise constraining the freeze-out conditions.
 Thus,  at present there exist many  different estimates of viscosity
coefficients, being  not consistent with each other and only indirectly
related to the observed flow effects. Thus no definite conclusion can
yet be done about temperature and density dependence  of the transport
coefficients in the QGP and hadron phases.

\section{Beyond quasiparticle approximation}\label{Res}

\subsection{Generalization to finite particle mass-widths}

Expressions (\ref{shear}), (\ref{bulk})  for the shear and
bulk viscosities  were derived within the quasiparticle
approximation, {\it i.e.}, the particle mass-width $\Gamma
=-2\mbox{Im}\Sigma^R$ was put zero in the retarded Green function
$G^R$. Here $\Sigma^R$ is the retarded self-energy. The spectral
density ($\widehat{A}=-2\mbox{Im}\widehat{G}^R$) is then expressed
through the $\delta$-function.

Let us start with the consideration of a fermion  spin $1/2$
resonance, "f". The spectral function satisfies the sum rule:
 \be\label{fsum-r}
\frac{1}{4} \mbox{Tr} \int^{\infty}_0 \gamma_0
\left[\widehat{A}^{\rm f}_{ (+)} (p_0 ,\vec{p})+ \widehat{A}^{\rm
f}_{ (-)} (p_0 ,-\vec{p})\right] \frac{d
  p_0}{2\pi}=2,\,\,\,
 \ee
where $\gamma_0$ is the corresponding Dirac matrix, and subscripts
$(\pm )$ specify particle and antiparticle terms.
 The trace is taken over spin degrees of freedom.
Simplifying the spin structure we introduce the spectral density
 \be\label{an}
 A^{\rm f}=\frac{1}{4} \mbox{Tr}\{ \ \gamma_0 \widehat{A}^{\rm f}_{
(+)}\} .
 \ee
Consider the dilute matter assuming that the spectral function
depends only on the  variable $s=p_0^2 -\vec{p}^{\,2}
>0$, see \cite{Voskr}. To do the problem  tractable,
instead of solving a complete set of the Dyson equations, we will
use a simplified phenomenological expression \cite{KTV08}:
 \be\label{fenomen}
  A^{\rm f} =\bar{A}^{\rm f}2p_0 =\frac{\xi \
 2p_0 \ [\bar{\Gamma}^{\rm f}
(s)+\delta] }{(s -m^{\rm part *2}_{\rm f})^2 +[\bar{\Gamma}^{\rm
f} (s)+\delta]^2 /4}, \quad \xi =const,\,
 \ee
with  $\delta \rightarrow +0$ introduced to easier perform the
quasiparticle limit. The value $\bar{\Gamma}^{\rm f} (s)$ is the
width. Phenomenological expression for its $s$-dependence is
presented in Appendix B.

 For charged bosons  the spectral function has the
form:\footnote{As before, by the charge we mean any conserved
quantity like electric charge, strangeness, etc.}
 \be
 \label{fenomen1}
 A_{\rm bos}=\frac{\xi \
 [\Gamma_{\rm bos} (s)+\delta] }{(s -(m^{{\rm part}*}_{\rm bos})^2)^2
 +[\Gamma_{\rm bos}(s)+\delta]^2 /4 }.
 \ee
Replacing  $\Gamma_{\rm bos} (s)$ with $\bar{\Gamma} (s)$ and
$A_{\rm bos}$ with $\bar{A}$, we may  use Eq. (\ref{widthf1}) of
Appendix B as a phenomenological parameterization of the width.
The quantity $\xi$ is  introduced to fulfill the sum-rule in the
form
 \be \int_{0}^{\infty}ds \frac{\bar{A}^2_{\rm f} (s)
\bar{\Gamma}_{\rm f} (s)}{2}=\int_{0}^{\infty}ds \frac{{A}^2_{\rm
bos} (s) {\Gamma}_{\rm bos} (s)}{2}=1.
 \ee

 We continue to study dilute matter assuming that the
spectral function depends only on the $s$-variable. Only in this
case one may consider a single spectral function for vector
mesons, like $\om$ and $\rho$, whereas in the general case one
should introduce transversal   and longitudinal components.   Now
we conjecture expression for the shear
viscosity which generalizes quasiparticle expression obtained
above to the case of the finite mass-width
\begin{align}\label{shearA}
\eta&=\frac{1 }{15T}\sum_{a}\int g_a \frac{d^3
p_a}{(2\pi)^4}\int_{0}^{\infty}ds \frac{\bar{A}^2_a (s)
\bar{\Gamma}_a (s)}{2}\,\tau_a \frac{\vec{p}^{\,4}_a}{E^2_a}\,
F^{\rm eq}_a(p_0) (1\pm F^{\rm eq}_a (p_0) ).
\end{align}
Here we have put $\tau_a =2p_0 /\bar{\Gamma}_a (s)$,  $E_a
=\sqrt{(m^{\rm part *}_{a})^2 +\vec{p}^2}$,
 \be\label{4eq}
 F^{\rm
eq}_a (p_0)=\frac{1}{e^{(p_0 -\mu_a^* )/T}\pm 1}. \ee
 Derivation of this expression will be given elsewhere.
 In the nonrelativistic approximation
Eq. (\ref{shearA}) coincides with the corresponding expression
derived in \cite{BS06}.

In the quasiparticle approximation
 \be\frac{\bar{A}^2_{\rm f} \bar{\Gamma}_{\rm f}}{2}\rightarrow
2\pi\delta(s -(m^{\rm part *}_{\rm f})^2 ), \quad \frac{{A}^2_{\rm
bos} {\Gamma}_{\rm bos}}{2}\rightarrow 2\pi\delta(s -(m^{\rm part
*}_{\rm bos})^2 ),
 \ee
and we return to Eqs. (\ref{shear}) -- (\ref{bulk}) with $\tau_a
=2E_a /\bar{\Gamma}_a (E_a)$.  Here the relaxation time for the
$a$-quasiparticle is expressed through the width, rather than the
cross-section. Bearing in mind the optical theorem, these two
quantities should coincide within the quasiparticle approximation.
This statement however requires an additional check that we do not
perform in the given work, since our aim here is just to estimate
possible effects of finite widths on the viscosities. Moreover in
the case of a finite width the off-mass shell $\tau_a (p_0 , \vec{p})$
 enter our expressions.

\begin{figure}[thb]
 \hspace*{20mm}
\includegraphics[width=90mm,clip]{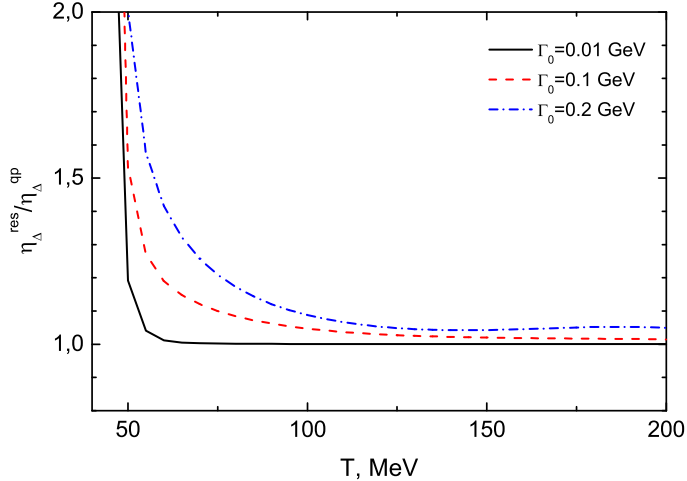}
\caption{  Temperature dependence of the ratio of the shear
viscosity  for the resonance calculated   in the constant width
approximation at different values of $\bar{\Gamma}=\Gamma_0 m_{\rm
res}$ to that computed in the quasiparticle approximation. In both
(resonance and quasiparticle) cases we use the same (for the given
$\Gamma_0$) energy-momentum independent values of the relaxation
times and $m_{\rm res}=m_{\Delta}$. Lines show different values of
the width.
 }
 \label{etaConst}
\end{figure}

 In Fig. \ref{etaConst}, we show the temperature dependence of the
ratio of the shear viscosity calculated for the  resonance
($m_{\rm res}=m_{\Delta}=$1.232 GeV) in the constant width
approximation ($\Gamma=\Gamma_0 m_{\rm res}$) with the constant
relaxation time to that computed in the quasiparticle
approximation with the same relaxation time. Realistic value of
the $\Delta$-resonance width in the resonance maximum is
$\Gamma^{\rm n.rel.} \simeq 120$ MeV that corresponds to the
choice $\Gamma_0 =0.24$ GeV.
 In order to
understand the dependence of the shear viscosity on the value of
the width we vary the value $\Gamma_0$.  We see that width effects
are quite important for the description of the resonance
characteristics at low temperatures $T\lsim 50\div 100$ MeV,
whereas at higher temperatures one may use  quasiparticle
approximation.
 At low temperatures contribution of
the resonances to the transport coefficients and to the
thermodynamical quantities is suppressed compared to the nucleon
and pion contributions. However, since even for very small width
(for $\Gamma_0 =0.01$ GeV corresponding to $\Gamma^{\rm n.rel.}
\sim 5$ MeV) the quasiparticle approximation for calculation of
the viscosity fails for $T<50$ MeV, and since nucleons  have some
width, we may conclude that calculations of the transport
coefficients at low temperatures should be performed with taking
into account width effects.

\begin{figure}[thb]
 \hspace*{20mm}
\includegraphics[width=90mm,clip]{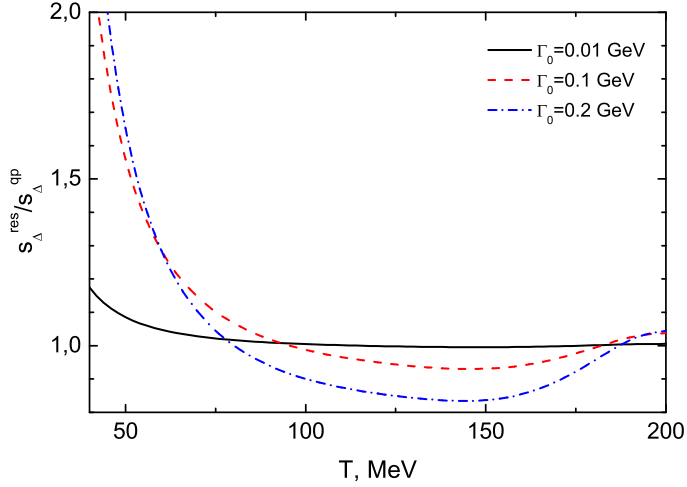}
\caption{  Temperature dependence of the ratio of the entropy
densities  for the resonance calculated  with  constant width and
in the quasiparticle approximation, $m_{\rm res}=m_{\Delta}$. The
curves are drown for different values of the width.}
 \label{sRes}
\end{figure}

\begin{figure}[thb]
 \hspace*{20mm}
\includegraphics[width=90mm,clip]{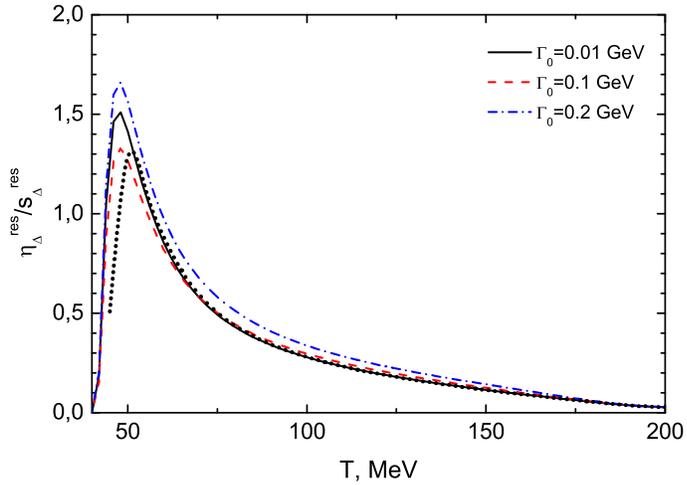}
\caption{  Temperature dependence of the specific shear viscosity
for the $\Delta$ resonance calculated   in the constant width
approximation, $m_{\rm res}=m_{\Delta}$. The quasiparticle result
is plotted by dots.
 }
 \label{etasD}
\end{figure}

 It was argued  in~\cite{Jakovac} that in a theory, where the lowest
energy excitations  (not quasiparticles) form a continuum, the
$\eta/s$ ratio has no lower bound in contrast with predictions of
the AdS/CFT correspondence. To check this statement we calculated
$\eta/s$ ratios. For the free resonance gas the entropy density is
given by \cite{Voskr,IKV00},
 \be s_{\rm f}=\sum_{a}g_a\int \frac{d^3 p}{(2\pi)^4}\int_{0}^{\infty}
 {ds}
 \bar{A}_{a}\sigma_{a}
 \ee
 where $\sigma_a =\mp (1\mp F_{a}^{\rm eq})\ln (1\mp
F_{a}^{\rm eq})-F_{a}^{\rm eq}\ln F_{a}^{\rm eq}$ and $F_{a}^{\rm
eq}$ is determined by Eq. (\ref{4eq}).

 In Fig. \ref{sRes} we present the ratio of the entropy
density computed with taking into account width effects to that
calculated in the quasiparticle approximation. As for the
viscosity, in case of small width (see solid curve for $\Gamma_0
=0.01$ GeV) the ratio $s^{\rm res}/s^{\rm qp}$ deviates from unity
only for low temperatures (at $T<50$ MeV). For a larger  width the
ratio begins to significantly differ from the unity in the whole
temperature interval.

In Fig. \ref{etasD} the temperature dependence of the $\eta^{\rm
res}/s^{\rm res}$ ratio is shown for $m_{\rm res}=m_{\Delta}$ and
for different values of $\Gamma_0$. We see that the dependence on
the value of the width is rather moderate. The ratio calculated in
the quasiparticle approximation behaves in a similar way but the
maximum position is slightly shifted toward a higher temperature.
For large temperatures the quasiparticle limit is achieved and,
thereby, within our model the $\eta/s$ ratio reaches the limit
$1/4\pi$ at $T\simeq T_c \sim 180$ MeV.

 \subsection{Mean-field bulk viscosity term}\label{mfv}

The sources of the shear and bulk viscosities considered above are
associated with collisions between in-medium excitations
(quasiparticles and particles with widths).
Another source of the bulk viscosity may come from a dissipation
of the soft modes related to a slow dynamics of the mean fields
provided
 the system is located in the vicinity of the critical point
of a second order phase transition or at a weak first order phase
transition. Note that in our SHMC model the soft modes relate to
the hadron masses decreasing towards the critical point $T_c$.
Such a contribution was first introduced in \cite{ML37,LL06} and
then discussed  in \cite{PP06,KhT07,KKhT07}.
 Within the quasiparticle
approximation used above, the mean fields do not dissipate.
Thereby, there are no linear in $\partial_{\mu}$ terms in
(\ref{varenmom}) responsible for an additional bulk viscosity
term. In reality there may exist such  a dissipation. We
demonstrate possible effect of a dissipation on the bulk viscosity
in Appendix C. However we should note that considering the soft
mode viscosity regime one assumes that the time characterizing the
soft mode evolution is sufficiently long. Estimations ~\cite{SV09}
show that soft modes may not develop during the time, at which the
system trajectory in the course of the heavy ion collision passes
the vicinity of the critical point. Thus   the  particle
collisions look a more adequate source of the viscosity  in  heavy
ion collisions than a source  due to the soft mode evolution.
Therefore  in the given paper we focused  on discussion of  the
collisional source for the viscosities.

\section{Conclusions}\label{Concl}

 In this paper, we analyzed different approximations for  the
calculation of the shear and the bulk viscosities for the hadron
and the quark-gluon phases. General expressions for shear and bulk
viscosities are   derived in the relaxation-time
approximation for a system described by the quasipartcle
relativistic mean-field theory with the scaling  of hadron
masses and couplings (SHMC). The EoS of the SHMC model fairly well
reproduces global properties of hot and dense  hadron matter
including the temperature region near $T_c$   and the lattice
data for $T_c\lsim T\lsim 1.3 T_c$, provided all coupling
constants $g_{\sigma b}$ are strongly suppressed except for the
nucleons.

 We compared kinetic coefficients calculated in the  SHMC model
with those calculated in other models  of the hadron phase. At
$\mu_{\rm bar} =0$ a  general fall off of the specific shear
viscosity with temperature in hadronic phase is a common feature
for all models but quantitative values  are somewhat different
particularly showing different behavior in the temperature region
near the transition temperature $T_c$. The $\eta/s$ ratio in the
SHMC model is closer to that of the excluded volume
hadron-resonance gas model with the hard-core radius $r\simeq 0.7$
fm~\cite{GHM08}  and practically coincides  in the near-critical
region with a resonance gas model  result that includes Hagedorn
states~\cite{NHGS08}.

For $\mu_{\rm bar}\neq 0$ transport coefficients qualitatively
behave similarly to the case of baryonless matter reaching the
lower AdS/CFT bound for the specific shear viscosity
$\eta/s=1/4\pi$ with increase of  $T$.   For the higher baryon
density this limit is reached at  smaller $T$.  In particular,
with increasing freeze-out temperature $T_{\rm fr}$ (for central
Au+Au collisions), the $\eta/s$ ratio undergoes a monotonous
decrease approaching values close to the AdS/CFT  bound at $T\sim
T_c $,\footnote{ We remind  that there exist rather general
arguments that $\eta/s$ should have a minimum at the QCD phase
transition critical point  similar to that exists for helium,
nitrogen and water, see \cite{CKML06}.} while the $\zeta/s$ ratio
exhibits a maximum at $T_{\rm fr}\sim$85 MeV. In a broad
temperature interval the $\eta/s$ and $\zeta/s$ ratios are not
small and viscous effects can be noticeable. The viscosity values
at the freeze-out can be transformed into dependence on the
colliding energy $\sqrt{s_{NN}}$ (for central Au+Au collisions).
When the collision energy decreases, the $\eta/s$ goes up.  The
high-energy flattening of the $\sqrt{s_{NN}}$ dependence occurs at
quite low $\eta/s<0.2$. It implies that a small value of $\eta/s$
required for explaining a large elliptic flow observed at RHIC
could be reached in the hadronic phase. This might be an important
observation which we have demonstrated within the SHMC and the
ideal gas  (IG) models with the same hadron set.
 The $\eta/s$ and $\zeta/s$ ratios at the freeze-out
curve calculated within the SHMC and IG models agree with each
other; however,  rough estimates \cite{IMRS09,DB08,Pal10} of
$\eta/s$ extracted from
 comparison of dynamical model calculations with experimental data overshoot
these theoretical expectations by a factor of 2-5.  Note that
rapid processes at the freeze out may result in an additional
increase of the viscosities and the entropy production. These
effects are not incorporated  directly in the present
consideration but the use of experimental freeze-out
temperature and chemical potential allows one to hope that
indirectly they are effectively included.

 Using  the Gibbs conditions we extended our approach to
higher temperatures combining the SHMC model description of the
hadron phase with that of the heavy quark bag (HQB) model for the
quark-gluon phase.  It was demonstrated that this two-phase
SHMC-HQB model is in reasonable agreement with QCD lattice data
 at $\mu_{\rm bar}=$0 and $\mu_{\rm bar}\ne$0  at all temperatures
except a vicinity of the critical point. The  finite-size and
nonequilibrium effects of the phase mixture,  possibility of a
rapid phase transition without a phase mixture may lead to
significant additional dissipation provided the system trajectory
passes the vicinity of the critical point. Moreover one might be
need to improve the model  in order to include possibility of the
crossover for $\mu_{\rm bar} =0$.   Not considered in
our work  these effects require a special careful analysis.

 In the two-phase SHMC-HQB  model with the first order phase
transition there
appears a jump in both  $\eta/s$ and $\zeta/s$ at the critical
temperature at  any $\mu_{\rm bar}$. For $T>T_c$ at $\mu_{\rm
bar} =0$, $\eta/s$ grows with increase of the temperature,
whereas $\zeta/s$ approximately stays constant.
 The magnitude of both
ratios differs significantly from those estimated in the NJL model
\cite{SR08,SR08c}. Comparison of our approach with  the NJL model
\cite{SR08,SR08c} demonstrates that behavior of transport
coefficients near the critical point depends on the type of the
phase transition.

It is worthy to note that the $v_2$ analysis~\cite{Cha09}
indicates different values of $\eta/s$ for peripheral and central
collisions. Therefore,  it would be interesting to perform
hydrodynamical calculations using the $T-\mu_{\rm bar}$ dependent
transport coefficients rather than constant ones. Necessity of
such an approach was recently emphasized in~\cite{Cha09}. In
subsequent works we plan to use the SHMC model EoS with the
derived transport coefficients for this purpose.

 We also estimated influence of the finite width effects on
the transport coefficients.  For this aim we evaluated the shear
viscosity and the entropy density at non-zero width. Our findings
show that width effects are important for low temperatures (for
$T\lsim (50\div 100)$ MeV), whereas for higher temperatures one
may use the quasiparticle approximation.

 As was demonstrated in ~\cite{KhT07,KKhT07}
the $\zeta/s$ ratio may have a  sharp maximum at the critical
point. This statement is based on the assumption of the existence
of soft collective modes at temperatures near $T_c$.
The enhancement of the ratio $\zeta/s$ near the critical point
could have  interesting
consequences~\cite{DG85,HK85,J95,ADM06,TTM08}, provided the system
evolves very slowly. However it is unlikely that in the course of
relativistic heavy ion collisions the system spends enough time in
the vicinity of the critical point to develop slow soft collective
modes, see ~\cite{SV09,Zeldovich}. Moreover effective hadron
masses responsible for the softness of the modes, being calculated
within our SHMC model,  although decrease
 toward the critical point do not drop to
zero. Thus the quasiparticle-like collisional viscosity
estimates look more relevant than those follow from the
consideration of the slow soft mode dynamics.  One should remind
once more that other  sources for an increase in the bulk viscosity
may exist, as discussed in \cite{PP06}.
Therefore further studies of the given problem are still required.

\vspace*{5mm} {\bf Acknowledgements} \vspace*{5mm}

We are grateful to  K.K.~Gudima, Y.B.~Ivanov, Y.L.~Kalinovsky,
E.E.~Kolomeitsev, K.~Redlich  and V.V.~Skokov for numerous
discussions and valuable remarks. We are thankful to K.~Redlich
for providing tables for values of the relaxation time calculated
within the NJL model. This work was supported in part by the
 the DFG grants 436 RUS 113/558/0-3
and WA 431/8-1,   Ukrainian-RFBR grant N 09-02-90423-Ukr-f-a and
the Heisenberg-Landau grant.

\vspace*{5mm}

{\bf{Appendix A. Necessary formulas for derivation of transport
coefficients.}}

Here to take derivatives in (\ref{deltaF})  we follow the
line of Refs. \cite{ADM06,SR08}. The only difference is
that the strangeness conservation is additionally incorporated.

The energy density $\epsilon$ and the baryon and strangeness
charge density ($n_{\rm bar}$ and $n_{\rm str}$) conservations can
be expressed as
\begin{align}
\label{eps} \frac{\pd\epsilon}{\pd t}&=-(\epsilon+P) {\vec{\nabla}}\cdot
\vec u=-\left(T\frac{\pd P}{\pd T}+\mu_{\rm bar}\frac{\pd
P}{\pd\mu_{\rm bar} }+\mu_{\rm str} \frac{\pd P}{\pd\mu_{\rm
str}}\right){\vec{\nabla}}\cdot \vec u ,\\ \label{nb}
 \frac{\pd n_{\rm bar}}{\pd
t}&=-n_{\rm bar} {\vec{\nabla}}\cdot \vec u=-\frac{\pd P}{\pd\mu_{\rm
bar}}{\vec{\nabla}}\cdot  \vec u ,\\ \frac{\pd n_{\rm str}}{\pd
t}&=-n_{\rm str} {\vec{\nabla}}\cdot \vec u=-\frac{\pd P}{\pd\mu_{\rm
str} } {\vec{\nabla}}\cdot \vec u~. \label{ns}
\end{align}
Here we set $\vec{u}=0$ keeping only derivative terms of
$\vec{u}$. The pressure is expressed in terms of $T,\mu$
variables. Further expressing $P$ in terms of $\epsilon, n$
variables and using Eqs. (\ref{eps})--(\ref{ns}) we find
\begin{align}
\frac{\pd P[\epsilon ,n]}{\pd t}&=\frac{\pd P}{\pd
\epsilon}\frac{\pd \epsilon}{\pd t}+\frac{\pd P}{\pd n_{\rm
bar}}\frac{\pd n_{\rm bar}}{\pd t}+\frac{\pd P}{\pd n_{\rm
str}}\frac{\pd n_{\rm str}}{\pd t}\nl&= -\left[\frac{\pd P}{\pd
\epsilon}\left(T\frac{\pd P}{\pd T}+\mu_{\rm bar}\frac{\pd
P}{\pd\mu_{\rm bar}}+\mu_{\rm str}\frac{\pd P}{\pd\mu_{\rm
str}}\right)\right.\nl&\left.\quad+\frac{\pd P}{\pd n_{\rm
bar}}\frac{\pd P}{\pd\mu_{\rm bar}}+\frac{\pd P}{\pd n_{\rm
str}}\frac{\pd P}{\pd\mu_{\rm str}}\right] {\vec{\nabla}}\cdot \vec u .
\end{align}
On the other hand
 \be
\frac{\pd P[T,\mu]}{\pd t}= \frac{\pd P}{\pd
 T}\frac{\pd T}{\pd t}+\frac{\pd P}{\pd \mu_{\rm bar}}\frac{\pd
\mu_{\rm bar}}{\pd t}+\frac{\pd P}{\pd \mu_{\rm str}}\frac{\pd
\mu_{\rm str} }{\pd t}
 \ee
 and thus
\begin{align}\label{Tt}
\frac{\pd T}{\pd t}&= -T\left(\frac{\pd P}{\pd
\epsilon}\right)_{n_{\rm bar},n_{\rm str}} {\vec{\nabla}}\cdot \vec u
,\\ \frac{\pd \mu_{\rm bar}}{\pd t}&= -\left[\mu_{\rm
bar}\left(\frac{\pd P}{\pd \epsilon}\right)_{n_{\rm bar},n_{\rm
str} }+\left(\frac{\pd P}{\pd n_{\rm bar}}\right)_{\epsilon,n_{\rm
str} }\right] {\vec{\nabla}}\cdot \vec u \label{mbart},\\
 \frac{\pd
\mu_{\rm str}}{\pd t}&= -\left[\mu_{\rm str}\left(\frac{\pd P}{\pd
\epsilon}\right)_{n_{\rm bar},n_{\rm str}}+\left(\frac{\pd P}{\pd
n_{\rm str}}\right)_{\epsilon,n_{\rm bar}}\right] {\vec{\nabla}}\cdot
\vec u .
\label{mstrt}
\end{align}
 We use these equations to construct variation of the
energy-momentum tensor $\delta T^{ij}$ in Eq. (\ref{deltaTij}).

{\bf{Appendix B. Phenomenological expression for the width of the
resonance.}}

The phenomenological expression for the $s$-dependence of the
width $\bar{\Gamma}^{\rm f} (s)$ can be easily recovered in the
near threshold region, see \cite{KTV08},
 \be
 \label{widthf1}
 \bar{\Gamma}^{\rm f} (s)&=&\Gamma_0 \ F(s) \ m^{\rm f}
\left(\frac{s^{1/2}-s_{\rm th}^{1/2}}{m^{\rm f}-s_{\rm th}^{1/2}}
\right)^{\alpha}\theta (s-s_{\rm th})~,\,\,\\
F&=&\frac{1}{1+[(s-s_{th})/s_0]^\beta}.\nonumber
 \ee
Here $\alpha =1/2$ for the s-wave resonance and $3/2$ for the
p-wave resonance, $m^{\rm f}$ is the fermion resonance mass in
vacuum, and
$s_{\rm th}$ is the threshold value of $s$; $\Gamma_0$ is the
constant of the dimensionality of the energy. In order to use this
expression outside the near threshold region one  introduces the
form-factor $F(s)$, the $s_0$ is the cut-off constant and the
power $\beta >1+\alpha/2$. The parameters
can be adjusted to satisfy experimental data. The factor $\xi$ is
introduced to fulfill the sum-rule:
 \be \int_{0}^{\infty}\frac{
\rmd s }{2\pi} \bar{A}_{\rm f} =1 ~.
 \ee

For charged bosons the spectral function follows the sum-rule,
 \be
\int_{0}^{\infty}\frac{ d s}{2\pi}A_{\rm bos}=1~.
 \ee

 {\bf{Appendix C. Contribution of soft collective modes to the bulk
 viscosity}}

As follows from the particle data, in vacuum the width of the
$\sigma$ meson is very large, $\Gamma_{\sigma} \sim 400$~MeV. In
hot baryonic matter the width can  additionally increase due to a
collisional broadening.
 Supposing $\sigma
=\sigma_{\rm eq}+\delta\sigma$ in the equation of motion for the
mean field, where $\sigma_{\rm eq}$ is the equilibrium value of
the mean field, we arrive at the equation for the fluctuating mean
field part $\delta\sigma$:
 \be\label{flS}
 (m^{*\rm part}_{\sigma}[T,\mu])^2\delta\sigma =
 -\Gamma_{\sigma}\frac{\partial \delta\sigma}{\partial t},
 \ee
 $1/\Gamma_{\sigma}$ is the relaxation time for the given process.
We suppressed the second space-time derivative terms but included
the first time-derivative dissipation term in this equation with a
 pre-factor $\Gamma_{\sigma}$. From Fig. \ref{s-vel_h} we see that
for the zero baryon chemical potential in the SHMC model with
suppressed couplings $g_{\sigma b}$, except for the nucleons,
which we use in the given paper, the effective mass $m^{*\rm
part}_{\sigma}$ significantly decreases at $T\sim T_c$.

Using smallness of the time-derivative term in (\ref{flS}) we find
 \be
 \delta\sigma =- \frac{\Gamma^{\rm eq}_{\sigma}}{(m^{*\rm
part}_{\sigma})_{\rm loc.eq}^2} \ \frac{d \sigma_{\rm
loc.eq}}{dt},
 \ee
 where $\sigma_{\rm loc.eq}$ is the mean field value at the
local equilibrium. On the other hand,
 \be
 \frac{d\sigma_{\rm loc.eq}}{d t}=
 \frac{d \sigma_{\rm loc.eq}}{d s} \ \frac{d s}{d t}
 =\frac{d \sigma_{\rm loc.eq}}{d s} \ s_{\rm loc.eq} \ \vec{\nabla}\cdot\vec{u},
 \ee
 where  we used approximate conservation of the entropy (up to terms
proportional to transport coefficients  assumed to be small).
Since the bulk viscosity contribution of the mean field is related
to the pressure as $\zeta_{\rm MF} \ \vec{\nabla}\cdot \vec{u}=
-\delta P_{\rm MF}$, we find
 \be
 \zeta_{\rm MF}=-\left(\frac{\partial
P_{\rm
MF}}{\partial\sigma}\right)_{\epsilon}\frac{\delta\sigma}{\vec{\nabla}
\cdot\vec{u}}  =\left(\frac{\partial
 P_{\rm MF}}{\partial\sigma}\right)_{\epsilon} \ \left[\frac{\Gamma}{(m^{*\rm
part}_{\sigma})^2}\frac{d\sigma}{ds}s\right]_{\rm loc.eq}~,
 \ee
and $[\frac{d\sigma}{ds}]_{\rm
loc.eq}=[\frac{d\sigma}{dT}/\frac{ds}{dT}]_{\rm loc.eq}$ is taken
at the condition that equations of motion (\ref{extreme1}) are
 fulfilled.

 There exist arguments that the $\rho$-meson also becomes a broad
resonance at $T$ near $T_c$, see \cite{Hees}, and its effective
mass decreases with increase of $T$ towards $T_c$. The $\rho$-mean
field appears in isotopically asymmetric nuclear  matter. In this
case, there may appear an extra contribution  to the bulk
viscosity. A similar contribution from the fluctuation of the
$\om_0$ field is probably suppressed compared to that from the
$\sigma$, since $\Gamma_{\om}\ll \Gamma_{\sigma}$.

\end{document}